\documentclass[iop]{emulateapj}
\usepackage{float}
\usepackage{natbib}
\usepackage{color,soul}
\usepackage{adjustbox,lipsum}
\usepackage{array}
\usepackage{subfigure}
\usepackage{threeparttable}
\usepackage[normalem]{ulem}
\usepackage[dvipsnames]{xcolor} 

\shorttitle{Void Galaxies in ECO}
\shortauthors{Florez et al.}

\begin{document}


\title{Void Galaxies Follow a Distinct Evolutionary Path in the Environmental COntext Catalog}


\author{Jonathan Florez\altaffilmark{1}, Andreas A. Berlind\altaffilmark{2}, Sheila J. Kannappan\altaffilmark{3}, David V. Stark\altaffilmark{4}, Kathleen D. Eckert\altaffilmark{5}, Victor F. Calderon\altaffilmark{2},  Amanda J. Moffett\altaffilmark{6}, Duncan Campbell\altaffilmark{7}, Manodeep Sinha\altaffilmark{2,8,9}}

\altaffiltext{1}{Department of Astronomy, University of Texas at Austin, Austin, TX 78712, USA}
\altaffiltext{2}{Department of Physics and Astronomy, Vanderbilt University, Nashville, TN 37235, USA}
\altaffiltext{3}{Physics and Astronomy Department, University of North Carolina, Chapel Hill, NC 27516, USA}
\altaffiltext{4}{Department of Physics and Astronomy, Haverford College, Haverford, PA 19041, USA}
\altaffiltext{5}{Department of Physics and Astronomy, University of Pennsylvania, Philadelphia, PA 19104, USA}
\altaffiltext{6}{Department of Physics and Astronomy, University of North Georgia, Oakwood, GA 30566, USA}
\altaffiltext{7}{McWilliams Center for Cosmology, Department of Physics, Carnegie Mellon University, Pittsburgh, PA 15213, USA}
\altaffiltext{8}{Centre for Astrophysics and Supercomputing, Swinburne University of Technology, Hawthorn, VIC 3122, Australia}
\altaffiltext{9}{ARC Centre of Excellence for All Sky Astrophysics in 3 Dimensions (ASTRO 3D)}


\begin{abstract}
We measure the environmental dependence, where environment is defined by the distance to the third nearest neighbor, of multiple galaxy properties inside the Environmental COntext (ECO) catalog. We focus primarily on void galaxies, which we define as the $10 \%$ of galaxies having the lowest local density. We compare the properties of void and non-void galaxies: baryonic mass, color, fractional stellar mass growth rate (FSMGR), morphology, and gas-to-stellar-mass ratio (estimated from a combination of HI data and photometric gas fractions calibrated with the RESOLVE survey). Our void galaxies typically have lower baryonic masses than galaxies in denser environments, and they display the properties expected of a lower mass population: they have more late-types, are bluer, have higher FSMGR, and are more gas rich. We control for baryonic mass and investigate the extent to which void galaxies are different at fixed mass. Void galaxies are bluer, more gas-rich, and more star forming at fixed mass than non-void galaxies, which is a possible signature of galaxy assembly bias. Furthermore, we show that these trends persist even at fixed mass and morphology, and we find that voids host a distinct population of early-types that are bluer and more star-forming than the typical red and quenched early-types. In addition to these empirical observational results, we also present theoretical results from mock catalogs with built-in galaxy assembly bias. We show that a simple matching of galaxy properties to (sub)halo properties, such as mass and age, can recover the observed environmental trends in ECO galaxies.
\end{abstract}



\section{Introduction}
Galaxies that live in low-density and void-like regions are ideal for studies of galaxy evolution due to the pristine environment in which they grow. Mergers and other transformation processes that are common in clusters, such as ram-stripping, are far less present in low-density regions. By studying the galaxies that occupy these void-like regions, one can begin to uncover the careful balance between nature and nurture in the context of galaxy formation.

Cosmological simulations have shown that halos in voids are less massive and that halo structure is largely dependent on environment \citep{2005ApJ...634...51A,2006ApJ...652...71W,2007ApJ...657..664J,2007MNRAS.377L...5G,2007MNRAS.376..215B}.  \cite{2009MNRAS.394.1825F,2010MNRAS.401.2245F} found that the dark matter merger rate has a strong correlation with environmental density; mergers dominate dense regions while diffuse accretion dominates halo growth in voids, thus explaining why voids in the dark matter distribution tend to be occupied by low-mass halos. For a long time, voids in dark matter simulations struggled to reproduce the observed spatial distribution of voids in surveys \citep{2001ApJ...557..495P}.  \cite{2009ApJ...691..633T}, however, were able to demonstrate that void properties are well described in the context of $\Lambda$CDM by assuming that the luminosity of galaxies is a function of the host dark matter halo mass and is independent of the large-scale halo environment.  

Extensive research has been done on the environmental dependence of galaxy properties in dense regions and clusters. \cite{1974ApJ...194....1O} and \cite{1980ApJ...236..351D} were some of the first to show that there exists a morphology-density relation in the distribution of galaxies, especially at higher densities. Over a decade of research has shown that galaxies in high-density regions are much more luminous and redder than those in low-density environments \citep[e.g.,][]{2003AJ....125.1866B,2004ApJ...615L.101B,2004ApJ...601L..29H}. Thanks to the wealth of data that now exist due to the advent of large redshift surveys, it has become possible to study galaxies in lower density environments in great detail.  Research focused on the galaxy population in low-density regions has revealed that these environments are populated with blue-sequence late-type galaxies \citep[e.g.,][]{1999AJ....118.2561G,2000AJ....119...32G,2004ApJ...617...50R,2007ApJ...658..898P,2009MNRAS.393.1324B} that are more star-forming and gas-rich than galaxies living in denser environments \citep[e.g.,][]{2005ApJ...624..571R,2007ApJ...654..702M}. 

Although the distributions of observable galaxy properties are largely dependent on mass and luminosity, current studies conflict over whether or not these properties depend on environment at fixed mass or luminosity. \cite{2004ApJ...615L.101B} proposed that the star formation of a galaxy is dependent on intrinsic properties, such as luminosity, and is independent of environment. \cite{2006MNRAS.372.1710P} found that the void galaxy property distributions are very similar to non-void galaxies and that there are no strong systematic differences in specific star formation rates between void and non-void galaxies. Similarly, \cite{2007ApJ...658..898P} found that variations between galaxy properties with different environments were almost entirely due to differences in the luminosity and morphology distributions. More recent work done by \cite{2014MNRAS.445.4045R}, however, presents evidence of void galaxies forming stars more efficiently than a control sample of the same stellar mass distribution. They show that larger voids have enhanced star formation activity in the shells surrounding the voids when compared to the smaller voids; however, they interpret this result as possibly being due to differences in dynamical evolution experienced by different sized voids. Additionally, work done by \cite{2012MNRAS.426.3041H} has found that void galaxies are statistically bluer than galaxies of the same magnitude distribution found in higher density environments and that both late and early type galaxies are bluer in voids.

In a $\Lambda$CDM universe, it is well established that the clustering of dark matter halos is a strong function of halo mass \citep[e.g.][]{1996MNRAS.282..347M}.  However, the clustering of halos is also known to depend on other properties like formation history \citep{2005MNRAS.363L..66G} and correlated properties like concentration, spin, and measures of local environment \citep{2006ApJ...652...71W, 2007MNRAS.377L...5G, 2018MNRAS.474.5143M}. Generally, halos that assembled at later times are found to be less clustered than halos of equal mass that assembled at earlier epochs, with the difference in clustering typically being most significant in lower-mass halos. This effect is referred to as ``halo assembly bias".

Environment is found to have a particularly strong correlation with the formation time of halos: lower density environments are populated by halos that assembled at later times compared to halos of the same mass living in higher density environments \citep{2009MNRAS.394.1825F, 2010MNRAS.401.2245F}. However, it remains unclear, what effect, if any, halo assembly history has on galaxy properties at fixed halo mass; therefore, it is not clear if galaxies also reflect assembly bias, hereafter a phenomena referred to as ``galaxy assembly bias". Nevertheless, if the structural properties of halos are correlated with large-scale environment through halo assembly bias, then it seems at least plausible that the correlation that exists between galaxy properties and large-scale environment is also due in part to assembly bias.

Studies that have included galaxy assembly bias have generally concluded that galaxy assembly bias is required in order to explain the observed colors and star formation rates of galaxies in semi-analytic models \citep{2007MNRAS.374.1303C} and abundance matching \citep{2014MNRAS.443.3044Z, 2014MNRAS.444..729H}, although see \citet{2018MNRAS.476.1637Z} for a conflicting result.  There is even evidence for galaxy assembly bias in cosmological hydrodynamical simulations of galaxy formation where assembly bias is not explicitly modelled \citep{2015ApJ...812..104T}.

In this paper we aim to investigate whether void environment plays any role in galaxy formation beyond simply providing different mass and morphology distributions. We use an Nth nearest neighbor method to define our void sample and provide an argument for using this method over Voronoi tessellations when defining a low density sample. We examine the effect of void environment on color, star-formation, gas content, and morphology in observations and we compare to the effect of void environment on color, star-formation, and gas content in a theoretical simulation.

The work is structured as follows. In Section 2 we introduce the data sets, observable properties, and environment metrics that are used for the analysis presented in this work. In Section 3 we present our findings on the properties of galaxies in voids. In Section 4 we compare our results to simulations designed to test assembly bias. Finally in Section 5 we present our conclusions.

\section{Data and Methods} 
\subsection{ECO/RESOLVE} \label{eco_data}
The Environmental COntext (ECO) catalog, a survey of $\sim13,000$ ``local" galaxies with $z \leq 0.023$, is the sample considered for this work (see \cite{2015ApJ...812...89M} for original survey catalog description and \cite{2016ApJ...824..124E} for updated catalog photometry and galaxy properties). ECO was  constructed to act as a ``context survey" for the REsolved Spectroscopy Of a Local VolumE (RESOLVE) survey \citep[Kannappan et al. in prep;][]{2015ApJ...810..166E,2016ApJ...832..126S} in order to calibrate the effects of cosmic variance in large-scale structure. The goal of RESOLVE is to survey and study the stellar mass, baryonic mass, gas-to-stellar mass ratio, star-formation, and dynamical mass of galaxies in a volume-limited census in the local universe by combining photometric and spectroscopic data spanning multiple wavelength regimes. 

The ECO region spans a right ascension (RA) range between 130.05 to 237.45 degrees, a declination (DEC) range between -1 and 49.85 degrees, and a line-of-sight velocity range between 2,530 and 7,470 km/s. The region encompassed by ECO on-sky lies where the Updated Zwickly Catalog (UZC; \citealt{1999PASP..111..438F}) and Sloan Digital Sky Survey (SDSS; \citealt{2000AJ....120.1579Y}) redshifts overlap. The ECO region encloses the A-semester of the RESOLVE survey and incorporates redshifts from the RESOLVE, HyperLEDA \citep{2003A&A...412...45P}, GAMA \citep{2011MNRAS.413..971D}, 2dF \citep{2001MNRAS.328.1039C}, and 6dF \citep{2009MNRAS.399..683J} surveys. The majority of ECO galaxies are present in SDSS, while $\sim 7 \%$ of galaxies are added to the final catalog from other sources. We use a volume-limited sample that is complete in absolute magnitude ($12,698$ galaxies with $M_r < -17.33$) in order to define environment metrics, however, we use a volume-limited sample complete in baryonic mass for our analysis. The ECO galaxy sample is complete in stellar mass at $M_{*} > 10^{9.1} M_{\odot}$, having $7,448$ objects with stellar masses above this limit and complete in baryonic mass at $M_b > 10^{9.3} M_{\odot}$ \citep{2015ApJ...812...89M}, having $9,526$ objects with baryonic masses above this limit. We decide to use a baryonic-mass complete sample for this analysis because we are specifically interested in examining how environment affects the baryonic properties of void galaxies. Throughout the analysis we assume $H_0 = 70$ ${\rm km~s}^{-1} {\rm Mpc}^{-1}$, and we calculate distances using $D= cz/H_0$.

\subsection{Photometric Data \& Morphologies}
For the analysis presented in this paper, we wish to examine properties such as color, gas-to-stellar mass ratio, baryonic \& stellar mass, and fractional stellar mass growth rate (FSMGR). Galaxy properties and photometric data are described by \cite{2015ApJ...812...89M} and \cite{2016ApJ...824..124E} for the ECO catalog and by \cite{2015ApJ...810..166E} for the RESOLVE catalog. ECO stellar masses and colors are derived via spectral energy distribution (SED) fitting \citep[see][for description of SED fitting procedure]{2013ApJ...777...42K}. The colors we use here are corrected for internal extinction and have $k$-corrections implicitly included.

Morphological classifications for ECO are determined from a combination of by-eye classifications and a method that exploits the relation between magnitude and the $\mu_{\Delta}$ metric described in \cite{2013ApJ...777...42K} which combines the overall surface mass density with a surface mass density contrast. We refer the reader to \cite{2015ApJ...812...89M} for a full description of how morphologies are derived for all galaxies in ECO.

The FSMGR quantity, described in \cite{2013ApJ...777...42K}, is a property similar to the specific star formation rate (sSFR) of a galaxy. The FSMGR quantity is defined as FSMGR $\equiv M_{<\mathrm{Gyr}}/M_{>\mathrm{Gyr}}$, where $M_{<\mathrm{Gyr}}$ is the stellar mass of a galaxy that has formed in the previous Gyr, and $M_{>\mathrm{Gyr}}$ is the stellar mass that formed prior to the last Gyr. The sSFR of a galaxy, on the other hand, has the total stellar mass in the denominator meaning its numeric value can never exceed a value of 1, unlike FSMGR which can reach values of up to $\sim 10$ for highly star forming galaxies.

\subsection{Baryonic Properties of ECO Galaxies}
In ECO, HI data exist for $\sim 44 \%$ of all galaxies. HI data, and therefore HI gas masses, for ECO galaxies are obtained from the RESOLVE survey \citep{2016ApJ...832..126S} and the public Arecibo Legacy Fast ALFA (ALFALFA) $\alpha .40$ catalog \citep{2011AJ....142..170H}. We consider an HI detection to be robust if it is not confused (i.e., there are no nearby neighbors that could contaminate or confuse the detection) and if it has a signal-to-noise ratio greater than 4. For ECO galaxies lacking robust HI detections, we resort to indirect methods to obtain the gas masses. The gas-to-stellar mass ratio of an ECO galaxy can be determined from a photometric gas fraction (PGF) relation calibrated with the RESOLVE survey \cite[see][for full description of PGF technique]{2015ApJ...810..166E}. 

The PGF technique takes advantage of the relation between galaxy color and the $M_{\rm HI} / M_*$ gas fraction \citep{2004ApJ...611L..89K}. For a galaxy's given color, a probability distribution of gas-to-stellar mass ratios can be made. We use these probability distributions to determine the predicted gas and baryonic masses of galaxies. To derive the predicted HI mass, we implement the $(u-J)$ \& $b/a$ color calibration from \cite{2015ApJ...810..166E} and obtain gas-to-stellar mass ratio distributions for each galaxy. The median of a galaxy's gas-to-stellar mass ratio probability distribution is used to assign a value of $M_{\rm HI}/M_{*}$ to that galaxy, and the predicted HI mass is then derived from that value using the stellar mass from SED fitting. We multiply all HI gas masses, real or predicted, by a factor of 1.4 to account for Helium. Therefore, all values of gas mass presented in this paper are given by $M_{\rm gas} = 1.4 \times M_{\rm HI}$. 

\begin{figure}[hbtp!]
\begin{center}
\includegraphics[scale=0.53]{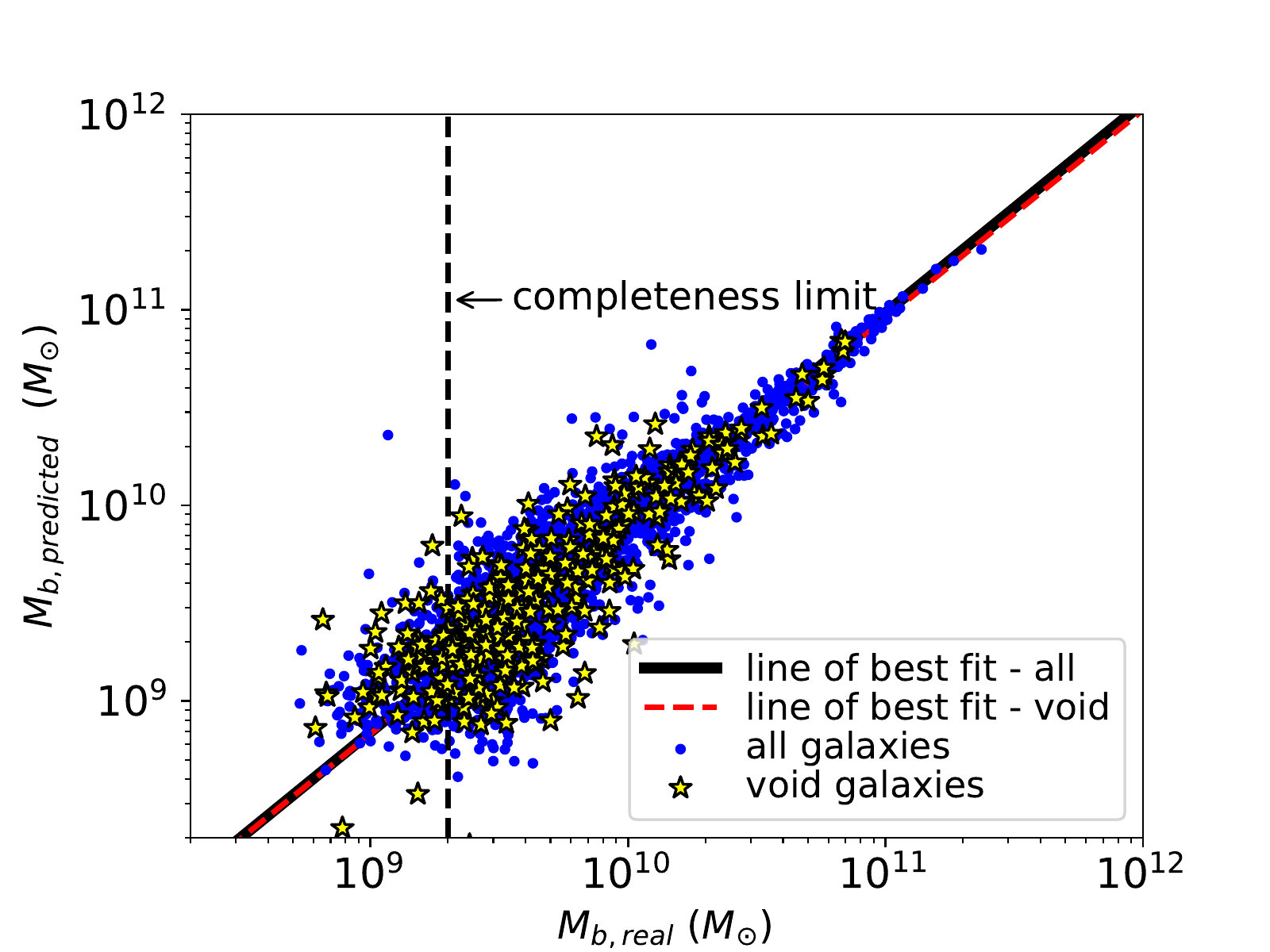}
\caption{The true baryonic mass for ECO galaxies with gas mass detections (x-axis) vs. the predicted baryonic mass derived from the PGF technique (y-axis). Void galaxies are shown as yellow stars while the rest of the sample is shown as blue circles. Also shown is the line-of-best fit, from a least squares regression, for void galaxies (red, dashed) and for all galaxies (black, solid). The line of best fit is only measured for galaxies with a baryonic mass above the completeness limit ($M_b > 10^{9.3} M_{\odot}$)}
\label{mbary_comparison}
\end{center}
\end{figure}

We compute baryonic masses by adding $M_*$ to $M_{\rm gas}$, where $M_{\rm gas}$ is computed from either the PGF technique or HI data depending on the strength of the HI detecion. We can compare a galaxy's predicted baryonic mass to its real baryonic mass if the galaxy has an HI detection either from the RESOLVE or ALFALFA surveys. The RESOLVE HI masses are obtained via 21cm data \citep{2016ApJ...832..126S}, and the RESOLVE-A region is only $78\%$ complete in HI when counting strong, unconfused detections (i.e. detections that have a signal-to-noise of $SN > 4$ and do not appear contaminated by nearby sources). We show the comparison between real and predicted baryonic masses in Figure \ref{mbary_comparison}. The error introduced by using a predicted baryonic mass diminishes towards higher masses, as the more massive galaxies are quenched and have baryonic masses that are dominated by the stellar mass. We measure the scatter to be $\sim 0.23$ dex at lower baryonic masses ($10^{9.3} M_{\odot} < M_{b,\rm{real}} < 10^{10.3} M_{\odot}$) and $\sim 0.09$ at higher baryonic masses ($M_{b, \rm{real}} > 10^{10.3} M_{\odot}$).

\subsection{Environment Metrics}
For the analysis described in this paper, we use an Nth nearest neighbor metric to measure the local galaxy density. We adopt $N=3$ when estimating densities with the Nth nearest neighbor method. We also performed the analysis using  $N = 3, 4, 5,$ and 10, however, there were no major variations in the results of the analysis.  In order to achieve robust statistics for our analysis, we assign void status to the $10\%$ of galaxies with lowest local density. We discuss this choice of method for selecting void galaxies in detail below.

\begin{figure*}[htbp!]
\begin{center}
\includegraphics[scale=0.7]{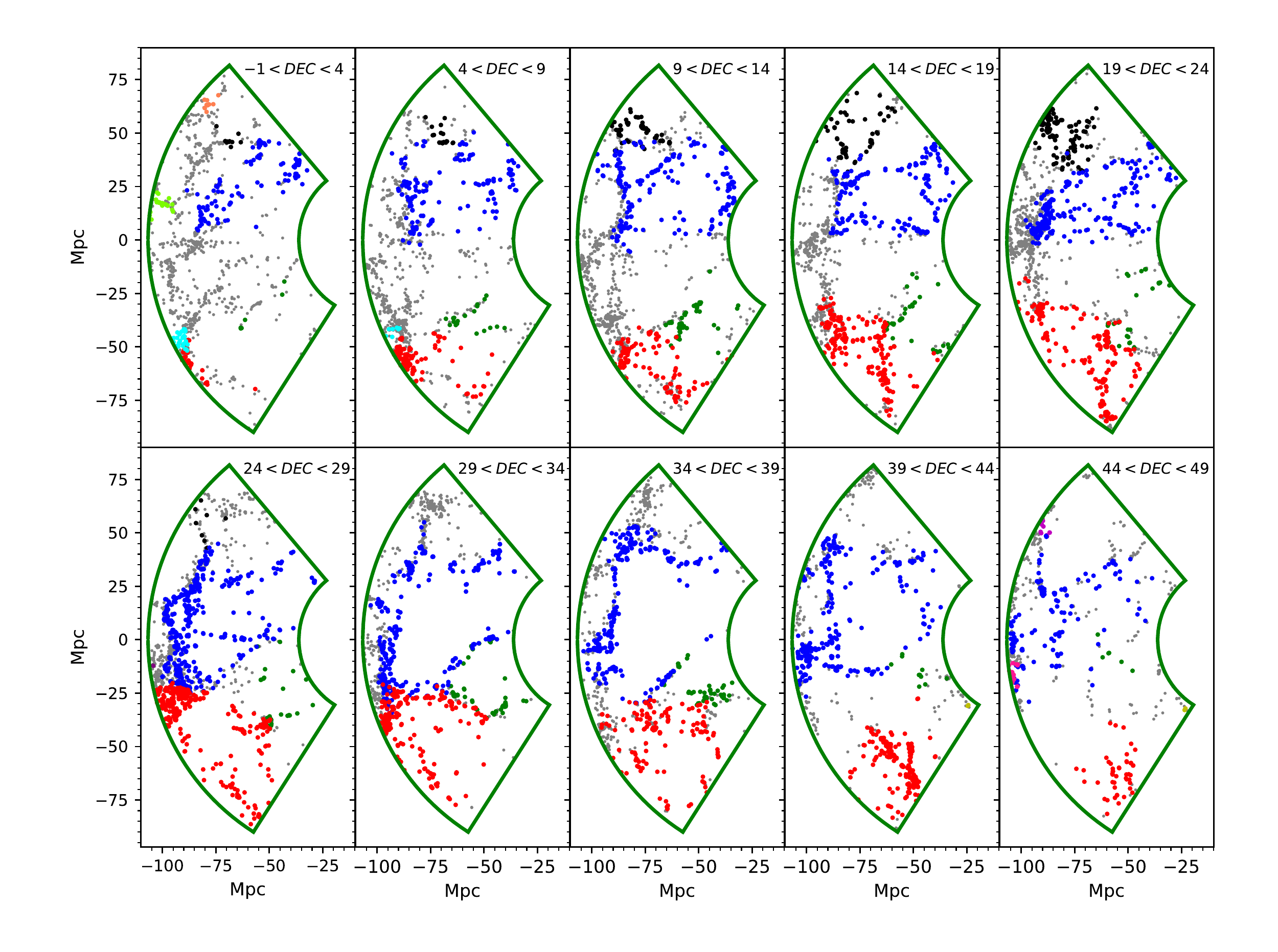}
\caption{The ECO sample in $5\,^{\circ}$ thick Dec slices increasing from the top left panel, in Cartesian coordinates. The top 10 percent most statistically significant voids identified by the ZOBOV algorithm are shown here, with each non-gray color representing a different void. All other galaxies are shown as gray points. It is clear that many galaxies that are classified as belonging to voids are actually in high density regions.}
\label{zobov_plots}
\end{center}
\end{figure*}

\subsubsection{Redshift Distortion Corrections}
We implement velocity corrections to the data in order to ensure that our density estimation is less sensitive to finger-of-god effects, such as lower density estimates for galaxies in large groups or cluster galaxies bleeding into voids. First, we run the \cite{2006ApJS..167....1B} friends-of-friends (FoF) group finder on the ECO data to determine group membership of the galaxies. The choice of FoF linking lengths are motivated by the work done in \cite{2016ApJ...824..124E} and \cite{2014MNRAS.440.1763D} who find optimal values of the linking lengths for the study of group-finding and environment. We determine the galaxy type (i.e. central or satellite) for each galaxy in the observation based on the galaxy's r-band absolute magnitude. The central galaxy is, by definition, the brightest galaxy in the galaxy group. We then ``collapse" the fingers of god by correcting the velocities of satellite galaxies. The correction process uses real-space and redshift-space versions of mock galaxy catalogs to recover the true radial distribution of galaxies in real space \citep[see][for complete description of velocity corrections]{2015ApJ...812...89M}.

We select void galaxy samples with and without finger-of-god corrections and find that 1,044 out of 1,270 galaxies (82\%) remain in both void galaxy samples applied to the ECO data. We have performed the analysis of this paper with the void galaxy sample selected from the ECO coordinates that do not have the finger-of-god corrections and find the results remain largely the same. Nonetheless, we keep the finger-of-god corrections in the analysis as this helps prevent cluster and group galaxies from falsely being classified as void galaxies.

\subsubsection{Void Classification}

\begin{figure}[hbtp!]
\begin{center}
\includegraphics[scale=0.53]{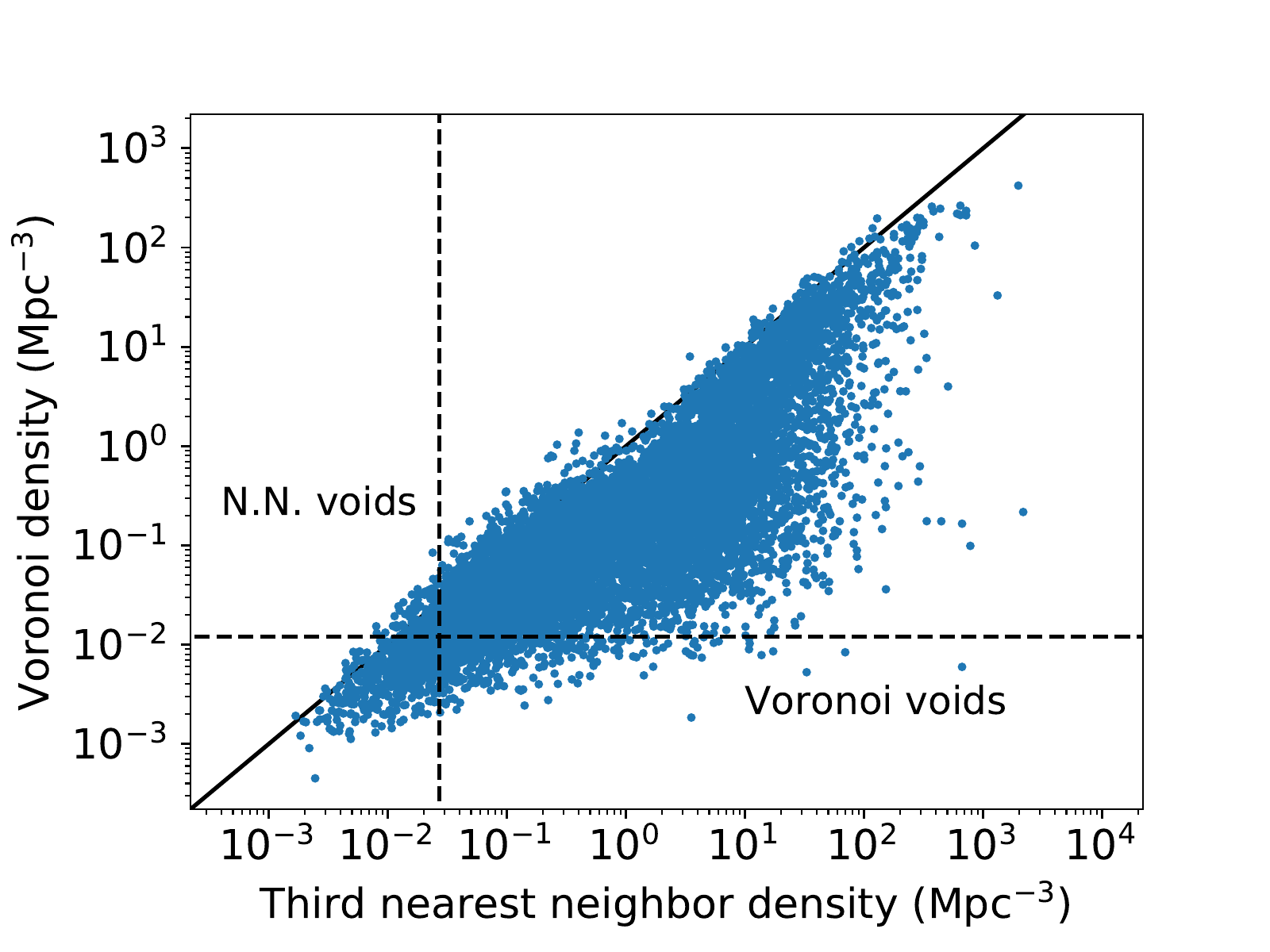}
\caption{Voronoi densities (y-axis) vs. third nearest neighbor densities (x-axis) of the ECO galaxy sample. The cut for void galaxies (indicated by the dashed lines) was arbitrarily chosen to categorize $\sim 10 \%$  of galaxies with the lowest densities as void galaxies by both methods.}
\label{density_comparison}
\end{center}
\end{figure}

\begin{figure*}[htbp!]
\begin{center}
\includegraphics[scale=0.7]{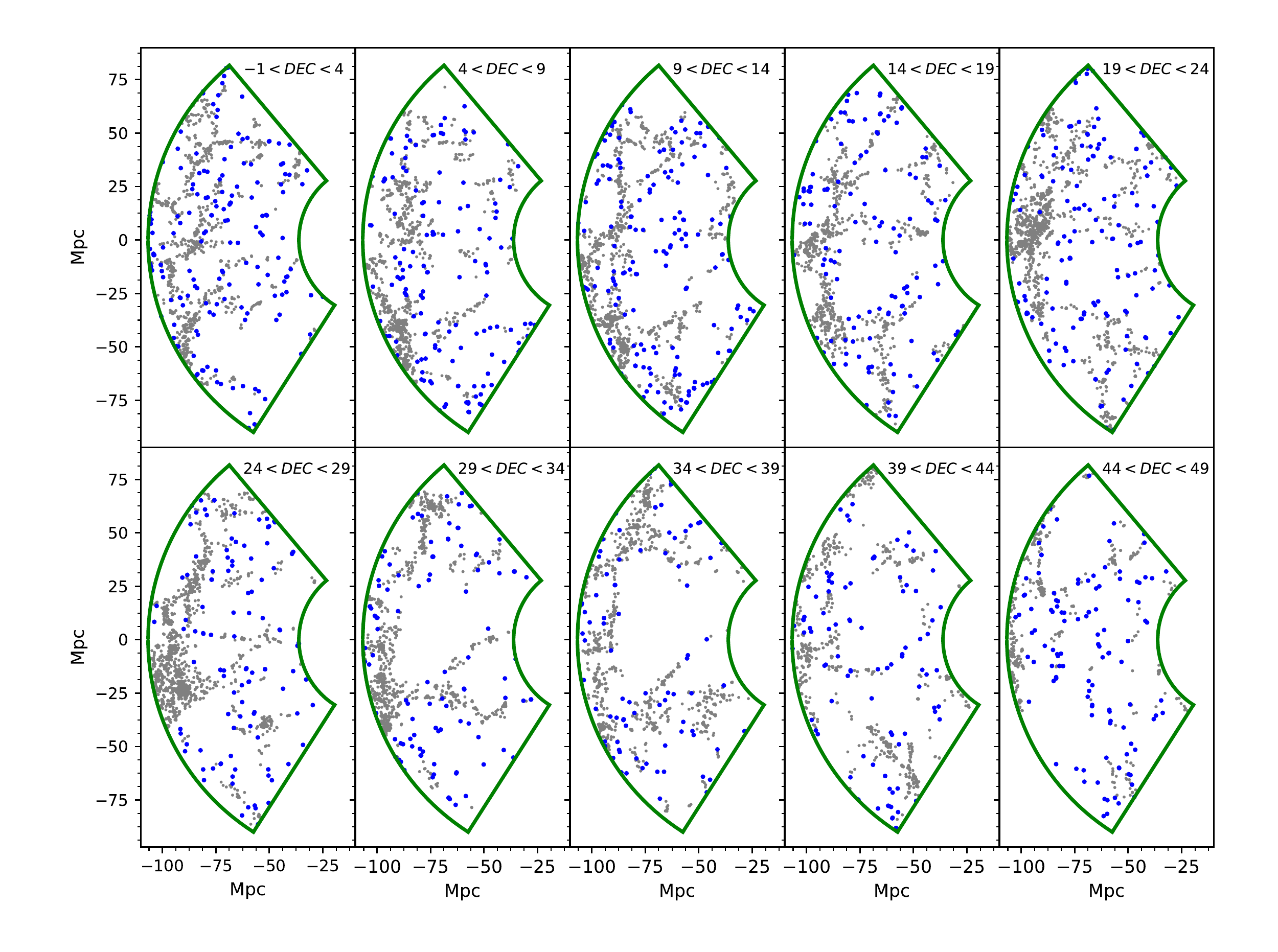}
\caption{Similar to Figure \ref{zobov_plots}, the ECO sample is shown here in $5\,^{\circ}$ thick Dec slices increasing from the top left panel. All void galaxies found by the Nth nearest neighbor algorithm are shown in blue, while non-void galaxies are shown as gray points.}
\label{low_density_plots}
\end{center}
\end{figure*}

In order to designate void status for a given galaxy, we need a way to quantify each galaxy's local density in ECO.  We initially experimented with the publicly available void-finding algorithm ZOBOV \citep{2008MNRAS.386.2101N} to find voids, but found that edge effects and having a low galaxy density often returned large spurious voids. The ZOBOV algorithm uses Voronoi tessellations to estimate local densities for galaxies and uses those densities to find density depressions in the sample volume. Because ZOBOV is designed to measure a void's shape and size, it includes galaxies living along the void border in its void definition. The problem with galaxies bordering the voids in ZOBOV is that those galaxies also border clusters and filaments, and often have relatively high local densities. We illustrate this in Figure \ref{zobov_plots}, which shows void galaxies found by ZOBOV. In order to exclude the void borders from our void definition, we decided to apply a density cut to the galaxies located inside the ZOBOV voids in order to create a better void sample. Since the majority of galaxies with low Voronoi densities also live in ZOBOV voids, the void sample remains essentially the same if one applies a density cut to the Voronoi densities of the entire sample of galaxies instead of just the ZOBOV void galaxies. For comparison purposes, we also estimate densities for galaxies using Nth nearest neighbor distances (adopting N=3) and, in both cases, we define void galaxies as the $10 \%$ of galaxies having the lowest density, which yields a statistically significant void sample. 

Figure \ref{density_comparison} directly compares the two methods for estimating density. The comparison reveals that some galaxies can have a low Voronoi density while having a high Nth nearest neighbor density, while the opposite does not hold for galaxies with low Nth nearest neighbor densities. A galaxy can be falsely classified as a void galaxy through the Voronoi tessellation method if that galaxy lives on the border of a cluster or filament because its Voronoi cell will extend into the void, thus having a large volume and a low corresponding density. In contrast, the Nth nearest neighbor density of such a border galaxy will be high. In Figures \ref{zobov_plots} and \ref{low_density_plots}, we show void galaxies chosen from both methods. In Figure \ref{zobov_plots}, we show void galaxies in the top 10 percent most statistically significant voids identified by ZOBOV. We clarify that this figure includes all galaxies in a given ZOBOV void region, as opposed to the void galaxies chosen by our lowest tenth percentile density cut. Even in the most statistically significant voids, it appears by-eye that many of the ZOBOV void galaxies live in high density regions. This is especially evident in the top right three panels of Figure \ref{zobov_plots}. In Figure \ref{low_density_plots} we show the entire void and non-void galaxy sample chosen by our Nth nearest neighbor metric. It is evident in Figure \ref{low_density_plots} that void galaxies chosen via our Nth nearest neighbor method appear to avoid regions of high density, as opposed to void galaxies identified by ZOBOV.

Given these results, we select the Nth nearest neighbor metric (with $N=3$) to define our void sample. We note that our void definition recovers galaxies that can either be completely isolated or exist in groups of two or three, and a few of the void galaxies even appear to be arranged in thin filamentary-like structures known as tendrils \citep[see][]{2013AJ....145..120B,2014MNRAS.440L.106A}. Although we choose our void definition based on $N=3$, the majority ($>90 \%$) of our void galaxies exist in FoF groups of $N=1$.

The threshold for selecting void galaxies for this analysis is exactly the lowest 10\% of galaxies ranked in density, selected from the entire sample of 12,698 ECO galaxies above the luminosity completeness limit ($M_r < -17.33$). This corresponds to a third nearest neighbor density $\rho_{\rm void}$ of less than $0.02$ galaxies Mpc$^{-3}$ for the void galaxy sample. The mean third nearest neighbor distance density $\bar{\rho}$ of the entire ECO volume is $0.76$ galaxies Mpc$^{-3}$, meaning the density contrast for the void sample is $\delta < -0.97$, where $\delta = \rho / \bar{\rho} - 1$.

We do not account for edge effects in this study, however, it is important to assess the impact this has on our analysis. To address this, we create a ``buffered" subsample of ECO data inside the RA, DEC, and line-of-sight velocity limits listed in Section \ref{eco_data} for all of ECO and rerun our void finding algorithm. The buffered region we create is the same as that from \cite{2015ApJ...812...89M}, which was specifically designed to mitigate edge effects that arise when computing galaxy environment metrics. The subsample of ECO data we create is 2.4 degrees smaller in RA, leaving a buffer of 1.2 degrees from each edge in RA, 2 degrees smaller in DEC, leaving a buffer of 1 degree from each edge in DEC, and 940 km/s shorter in line-of-sight velocity distances, corresponding to a line-of-sight velocity range of 3,000 to 7,000 km/s. These limits were chosen to ensure that the ECO subsample has a buffer region of at least 1 Mpc in all directions, with an additional allowance of $\sim 5.7$ Mpc in the line-of-sight velocity distances to compensate for group peculiar velocities that extend up to 400 km/s \citep{2015ApJ...812...89M}.

The subsample of ECO data we create has a total of 9,359 galaxies complete in absolute magnitude ($M_r < -17.33$) and a total of 936 void galaxies above this limit, corresponding to the lowest $10 \%$ of galaxy densities. Of these 936 void galaxies in the buffered subsample of ECO, 877 $(93.6 \%)$ are in the original (full) ECO sample. We find that the results of this paper do not change when we perform our analysis on the buffered subsample of void galaxies. We therefore decide to keep our original sample of 1,270 void galaxies, for this analysis.

\section{Results}
In Section \ref{void_properties} we examine the observable properties of the void sample and compare them to those of the non-void sample. In sections 3.2 \& 3.3 we investigate whether the different mass or morphology distributions of the samples drives the differences we see between void and non-void galaxies.

\subsection{Observed Properties of Void Galaxies} \label{void_properties}
\begin{figure}[htbp!]
\begin{center}
\centerline{\includegraphics[scale=0.55]{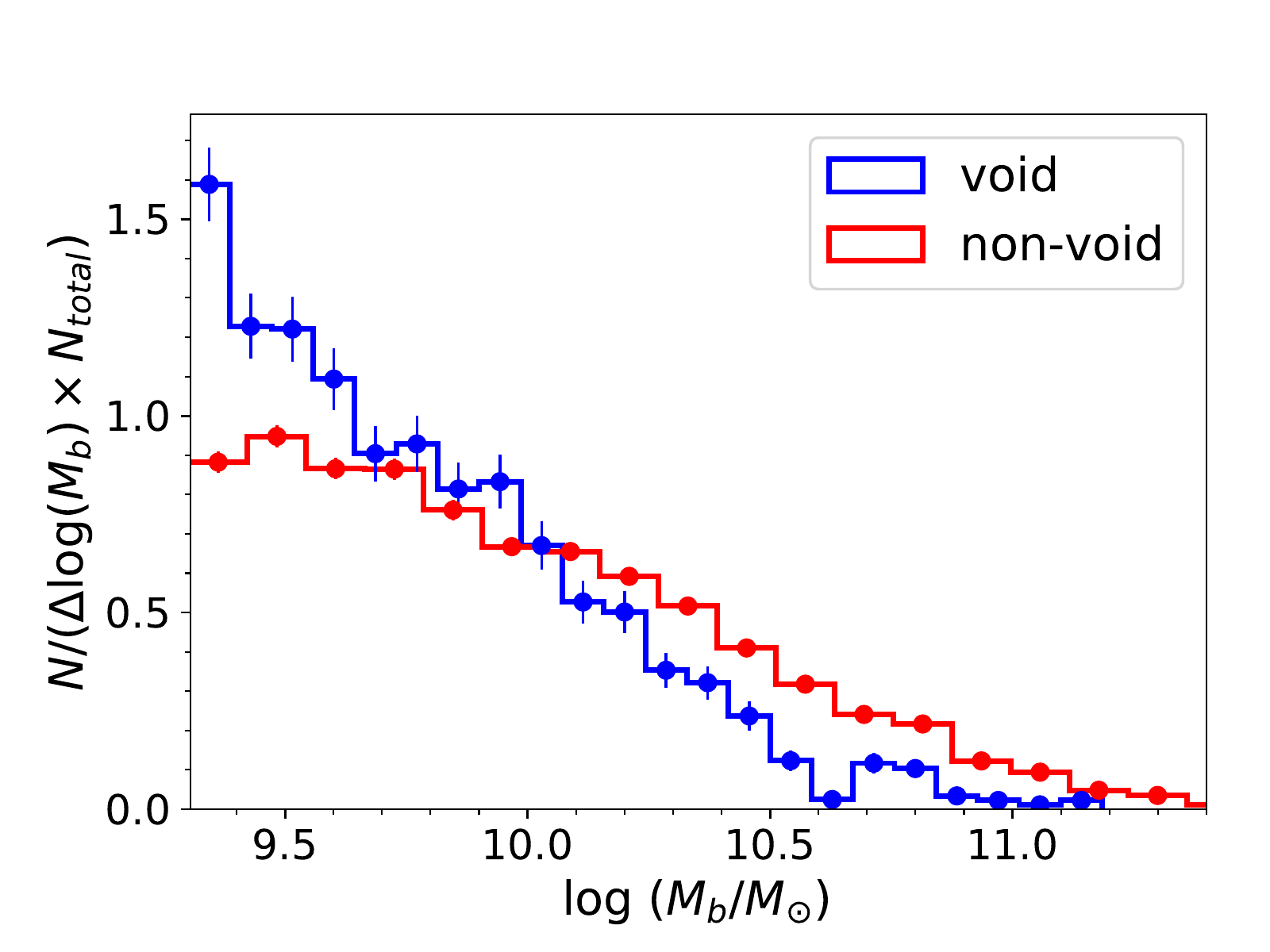}}
\caption{Normalized histogram of the baryonic mass distribution of galaxies in ECO for void galaxies (blue) and non-void galaxies (red) above the completeness limit with Poisson errors. }
\label{hist}
\end{center}
\end{figure}

In Figure \ref{hist} we show the baryonic mass distribution of the void and non-void galaxy sample. The samples have different mass distributions, with void galaxies being less massive on average than non-void galaxies. Our results show that void galaxies rarely have masses greater than $10^{10.6} M_{\odot}$. The void galaxy distribution has an excess in baryonic masses below $10^{9.8} M_{\odot}$ and a deficiency in masses above $\sim 10^{10} M_{\odot}$ when compared to the baryonic mass distribution of all ECO galaxies. While not shown here, we find a similar trend when comparing the stellar mass distributions of the void and non-void galaxy samples.

\begin{figure*}[htbp!]
\begin{center}
\includegraphics[scale=0.7]{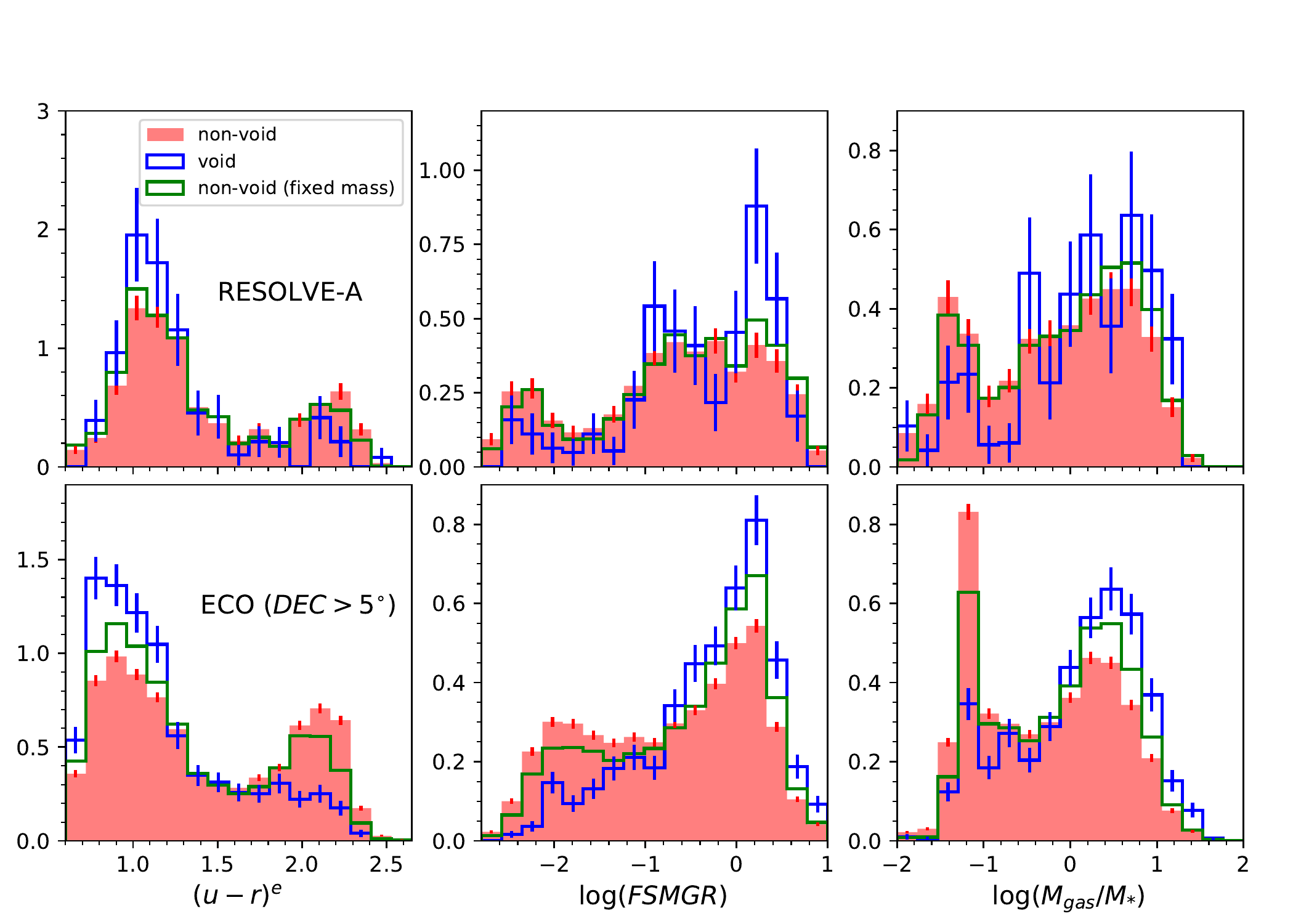}
\caption{This figure shows normalized histograms of color, FSMGR, and $M_{\rm gas}/M_{*}$ for void (blue) and non-void galaxies (red) in the RESOLVE-A semester (top) and ECO at $Dec. > 5^{\circ}$ (bottom). The green line represents a sample of non-void galaxies with a baryonic mass distribution that matches that of the void galaxy sample. In the top panels, values of $M_{\rm gas}$ come primarily from observations whereas in the bottom panels values of $M_{\rm gas}$ come from a combination of observed and predicted gas masses (see Section 2.3). All the data shown here is complete in baryonic mass above $M_b = 10^{9.3} M_{\odot}$.}
\label{new_pdf}
\end{center}
\end{figure*}

In Figure \ref{new_pdf} we examine normalized histograms of void and non-void galaxy properties in ECO and RESOLVE. We show the RESOLVE A semester catalog data in the top panels as it has real HI masses derived from 21cm data and compare that with ECO data in the bottom panels. We impose a cut of $Dec. > 5^{\circ}$ in ECO for this figure only so that the top panels contain and display independent information.

The histograms show that the ECO void galaxies are bluer, more star-forming, and more gas-rich than their non-void counterparts. RESOLVE shows qualitatively similar results, though they are much noisier due to the smaller size of that sample. We perform an Anderson-Darling (AD) $k$-sample test between the void and non-void galaxy property distributions for each panel shown in Figure \ref{new_pdf} using the {\tt scipy.stats} package from the SciPy library \citep{2020SciPy-NMeth}. We show the results of the test in Table \ref{AD1} and note that $p$-values are capped at values of $0.001$ and $0.25$, so $p$-values above (below) 0.25 (0.001) are only quoted as lower or upper limits. The results of the AD test show that the differences between void and non-void galaxy properties are statistically significant. The two left panels of Figure \ref{new_pdf} comparing $(u-r)^e$ colors between void and non-void galaxies can be compared to Figure 3 of \cite{2012MNRAS.426.3041H}, which shows that void galaxies are statistically bluer than non-void galaxies. These results are not surprising, since void galaxies are less massive and there are well established correlations between mass, color, and gas content, with lower mass galaxies having higher gas fractions \citep{2004ApJ...611L..89K}, higher specific star formation rates \citep{2005ApJ...621L..89B}, and bluer colors \citep{2004ApJ...600..681B, 1980ApJ...236..351D, 2012MNRAS.426.3041H}. The interesting question is whether these differences persist at fixed mass.

\begin{table}[H]
\begin{center}
\caption{Test statistic and $p$-values of the Anderson-Darling (AD) $k$-sample test performed between the void and non-void galaxy property distributions shown in Figure \ref{new_pdf}. Results are shown for both the ECO and RESOLVE-A semester samples.}
\begin{tabular}{| l c c |}
\hline
\multicolumn{3}{|c|}{RESOLVE-A}\\
\hline
galaxy property & AD statistic & $p$-value\\
\hline
$(u-r)^e$ & 5.95 & 0.002\\
$\log(\rm FSMGR)$ & 4.15 & 0.007\\
$\log(M_{\rm gas}/M_*)$ & 3.91 & 0.009\\
\hline
\multicolumn{3}{|c|}{ECO} \\
\hline
galaxy property & AD statistic & $p$-value\\
\hline
$(u-r)^e$ & 99.19 & $\leq 0.001$\\
$\log(\rm FSMGR)$ & 107.23 & $\leq 0.001$\\
$\log(M_{\rm gas}/M_*)$ & 104.58 & $\leq 0.001$\\
\hline 
\end{tabular}
\label{AD1}
\end{center}
\end{table}

\subsection{Are Void Galaxies Different at Fixed Mass?}

We move on to investigate how the properties of void galaxies differ from non-void galaxies when controlling for mass. Referring again to Figure \ref{new_pdf}, we compute histograms for a sample of non-void galaxies that have the same baryonic mass distribution as the void sample. To generate these histograms, we first determine the baryonic mass distributions of the void and non-void samples (shown in Figure \ref{hist}). We then randomly sample from the non-void galaxies with a probability that is proportional to the ratio of the mass distributions in order to build a sample of non-void galaxies with a mass distribution identical to that of the void galaxies. During the random sampling, we check to make sure that a galaxy isn't sampled twice. This process is then repeated 1000 times so that we end up with 1000 samples of non-void galaxies with a void-like mass distribution. The property distributions are all averaged together to generate the green histograms shown in Figure \ref{new_pdf}.

\begin{figure*}[htbp!]
\begin{center}
\centerline{\includegraphics[scale=0.74]{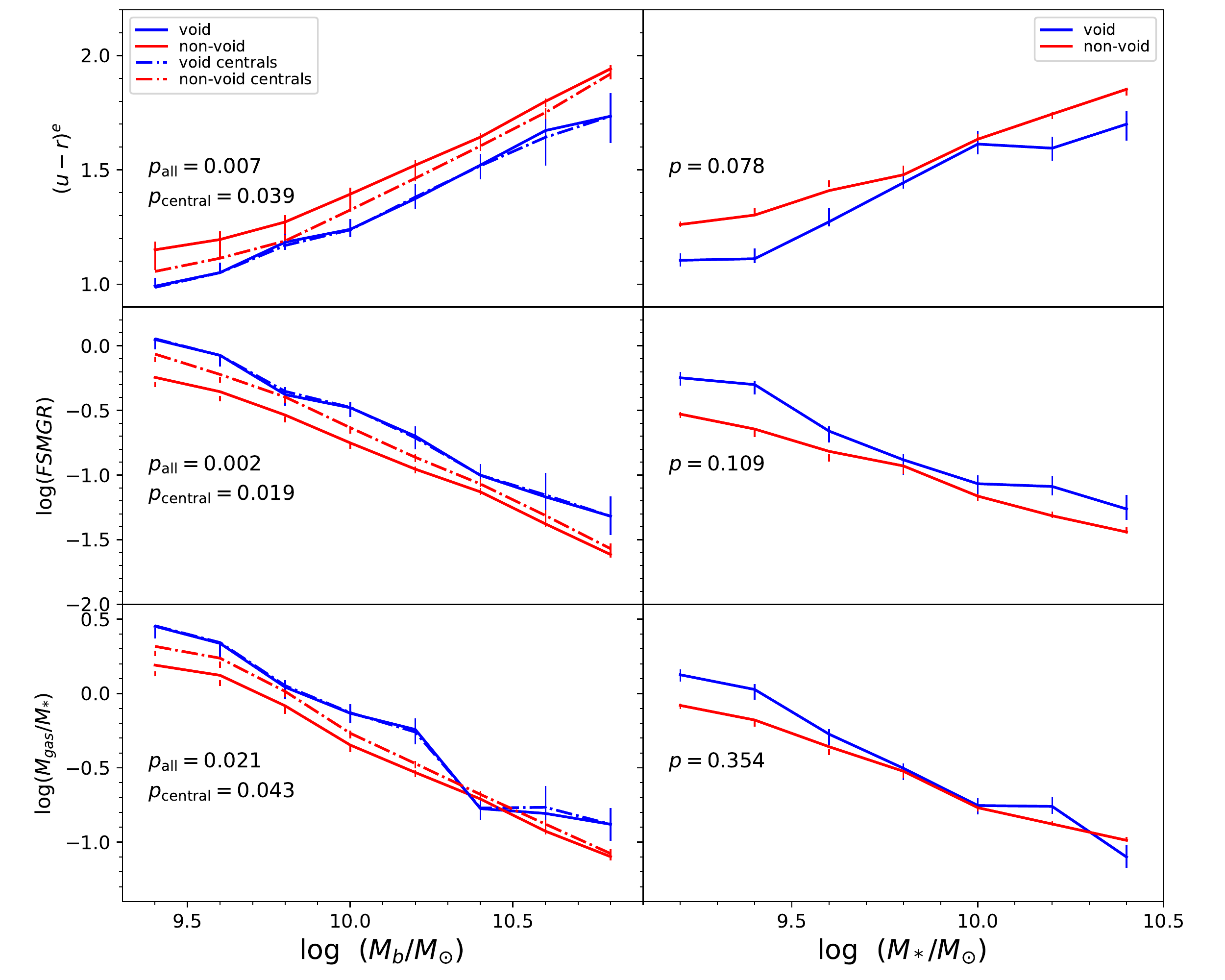}}
\caption{Left: For $(u-r)^e$ (top panels), $\log(FSMGR)$ (middle panels), and $\log(M_{\rm gas}/M_*)$ (bottom panels), we show the mean of each property in eight bins of baryonic mass for void galaxies (blue) and non-void galaxies (red) in intervals of $\Delta \log M_{b} = 0.2$. We show this for all void and non-void galaxies (solid) and for void and non-void galaxy group centrals (dot-dashed). Each data point is placed in the middle of the mass bin and error bars show $1 \sigma$ uncertainties derived from a bootstrap resampling inside each bin. The $p$-value of the $T_{n,3}$ test statistic is shown in each panel for all void and non-void galaxies and for void and non-void galaxy centrals (see text). Right: Similar to left panels, but galaxy properties are measured inside seven bins of stellar mass with $\Delta \log M_{*} = 0.2$. For clarity, we only show the data for all void and non-void galaxies in this plot (i.e., we do not make distinctions between satellites and centrals). We also show the $p$-value of the $T_{n,3}$ test statistic between void and non-void galaxies in each panel.}
\label{average_prop}
\end{center}
\end{figure*}

We find that differences between void and non-void galaxies persist when controlling for mass. The void galaxy sample is bluer, more gas-rich, and has higher FSMGR when compared to the mass-matched non-void galaxy sample, despite having the same mass distribution.  We note that the difference between the void (blue histograms) and mass-matched non-void (green histograms) galaxy samples is due to environment, while the difference between the mass-matched non-void (green histograms) and unaltered non-void (red histograms) samples is due to baryonic mass. We find that the mass-matched non-void histograms in ECO and RESOLVE-A decrease with respect to the void histograms, in fixed bins, by a factor of $\sim 0.2$ to $0.5$ at bluer colors, higher FSMGRs, and higher gas fractions. The mass-matched non-void histograms increase with respect to the void histograms, in fixed bins, by a factor of $\sim 1-3$ at redder colors, lower FSMGRs and lower gas fractions. As in Section \ref{void_properties}, we perform an AD $k$-sample test between the void and mass-matched non-void galaxy property distributions for each panel in Figure \ref{new_pdf} and present the results of the tests in Table \ref{AD2}. The results of the test show that the quantitative differences between void and non-void galaxy property distributions remain statistically significant in ECO, even when controlling for mass. These differences are less statistically significant in RESOLVE-A, though this could likely be an effect due to the low number of void galaxies $(< 100)$ in RESOLVE-A.

In Figure \ref{average_prop} we use an alternative method for controlling the mass distribution of void and non-void galaxies that lets us determine how galaxy properties depend on environment. We bin both samples in stellar and baryonic mass and compare the mean of $\log$(FSMGR), $(u-r)^e$, and $\log\left(M_{\rm gas}/M_*\right)$ for void and non-void galaxies in each bin. Error bars are $1\sigma$ confidence intervals made from a bootstrap resampling inside each bin. To compute the error from a bootstrap method, we resample (i.e., draw randomly from) the distribution inside each bin of mass $x$ times, where $x$ is the number of objects inside that bin, build a sample from the random draws and compute the mean. We draw objects with replacement so the same object in a given bin can be sampled more than once. We do this 1,000 times and make a distribution of the mean values inside each bin and take the 16th and 84th percentile of this distribution as the error. We find that void galaxies are bluer, more star-forming, and more gas-rich across almost all bins of baryonic mass. We find similar trends when controlling for the stellar mass, across all stellar mass bins.

In order to assess the statistical significance of the comparisons shown in Figure \ref{average_prop}, we perform a multivariate test for equal distributions based on nearest neighbors using the {\tt yaImpute} package \citep{2007Crookston} in R \citep{RMAN}. Following the procedure in Section 10.3 of \cite{2019.RIZZO}, we compute the third nearest neighbor test statistic, $T_{n,3}$, between void and non-void galaxies for the mass ranges shown in Figure \ref{average_prop} ($\log(M_b / M_{\odot}) = 9.3 - 10.9$ for left panels, $\log(M_* / M_{\odot}) = 9.1 - 10.5$ for right panels). We normalize the data, such that each variable has a value between 0 and 1, before computing $T_{n,3}$ and then perform a bootstrap of the $T_{n,3}$ test statistic to obtain a $p$-value. The $p$-value of each test is shown in the panels of Figure \ref{average_prop}.

We find higher $p$-values ($p > 0.05$) for comparisons between the void and non-void galaxy samples complete in stellar mass, indicating that the differences between the stellar-mass complete samples may not be as statistically significant as the differences between the baryonic-mass complete samples. The difference in sample sizes is important to note, however, as there are 895 total void galaxies being tested in the left panels of Figure \ref{average_prop}, 857 of which are centrals, while there are 522 total void galaxies being tested in the right panels of Figure \ref{average_prop}. Similarly, there are 8,288 non-void galaxies and 5,718 non-void galaxy centrals being tested in the left panels of Figure \ref{average_prop}, while there are 5,821 total non-void galaxies being tested in the right panels of Figure \ref{average_prop}. We note that these numbers vary from what is listed in Table \ref{cents} because we do not show the highest mass bins in Figure \ref{average_prop} due to very low $(< 10)$ sample sizes for the void galaxies. The difference in sample sizes inevitably leads to higher $p$-values for the void and non-void galaxy samples being tested in the right panels of Figure \ref{average_prop}.

\begin{table}[H]
\begin{center}
\caption{Test statistic and $p$-values of the AD $k$-sample test performed between the void and mass-matched non-void galaxy property distributions shown in Figure \ref{new_pdf}. Results are shown for both the ECO and RESOLVE-A semester samples.}
\begin{tabular}{| l c c |}
\hline
\multicolumn{3}{|c|}{RESOLVE-A}\\
\hline
galaxy property & AD statistic & $p$-value\\
\hline
$(u-r)^e$ & 3.02 & 0.019\\
$\log(\rm FSMGR)$ & 1.37 & 0.088\\
$\log(M_{\rm gas}/M_*)$ & 0.96 & 0.13\\
\hline
\multicolumn{3}{|c|}{ECO} \\
\hline
galaxy property & AD statistic & $p$-value\\
\hline
$(u-r)^e$ & 43.52 & $\leq 0.001$\\
$\log(\rm FSMGR)$ & 46.47 & $\leq 0.001$\\
$\log(M_{\rm gas}/M_*)$ & 41.85 & $\leq 0.001$\\
\hline 
\end{tabular}
\label{AD2}
\end{center}
\end{table}

Part of the differences observed between void and non-void galaxies at fixed mass can be attributed to contributions from satellite galaxies in the non-void populations. For this reason, we quantify how much the satellite population contributes to the differences we see between void and non-void galaxies at fixed mass in Figure \ref{average_prop}. On the left-hand panels in Figure \ref{average_prop}, we show the mean color, FSMGR, and gas fraction of group centrals for both void and non-void galaxies at fixed baryonic mass. We also compute the $T_{n,3}$ test statistic described above for void and non-void galaxy centrals and show the corresponding $p$-values in the left panels of Figure \ref{average_prop}. For reference, we show the number of centrals and satellites in the baryonic and stellar mass complete void and non-void galaxy samples in Table \ref{cstab}. The mean of void galaxy group centrals hardly changes, due to the fact that more than $90\%$ of all void galaxies exist inside N$=1$ groups. The mean of non-void centrals at fixed baryonic mass, however, appears bluer, more star-forming, and more gas-rich than the mean of all non-void galaxies, especially at lower masses. We note that the $p$-values of the $T_{n,3}$ test statistic for void and non-void galaxy centrals indicate that the differences between void and non-void galaxies remain statistically significant among the population of centrals. Lower mass non-void galaxies can exist as satellites in large groups and clusters, and those satellites can be expected to have redder colors, lower FSMGRs and gas fractions than galaxies outside of clusters of the same mass as they have been quenched through some of the environmental processes that take place in groups and clusters. Nevertheless, this satellite effect does not explain away all of the differences between void and non-void galaxies. Differences persist even when considering central galaxies alone.

\begin{table}[H] \label{cents}
\begin{center}
\caption{Number of satellites vs. centrals in the ECO baryonic-mass and stellar-mass complete void and non-void galaxy samples.}
\begin{tabular}{| c | c | c |}
\hline
Sample & $\log(M_b / M_{\odot}) > 9.3$ &  $\log(M_* / M_{\odot}) > 9.1$\\
\hline
\hline
Void centrals & 863 & 533 \\
Void satellites & 38 & 20\\
Non-void centrals & 5,999 & 4,622\\
Non-void satellites & 2,626 & 2,273\\
\hline
\end{tabular}
\label{cstab}
\end{center}
\end{table}

\begin{figure*}[htbp!]
\begin{center}
\centerline{\includegraphics[scale=0.65]{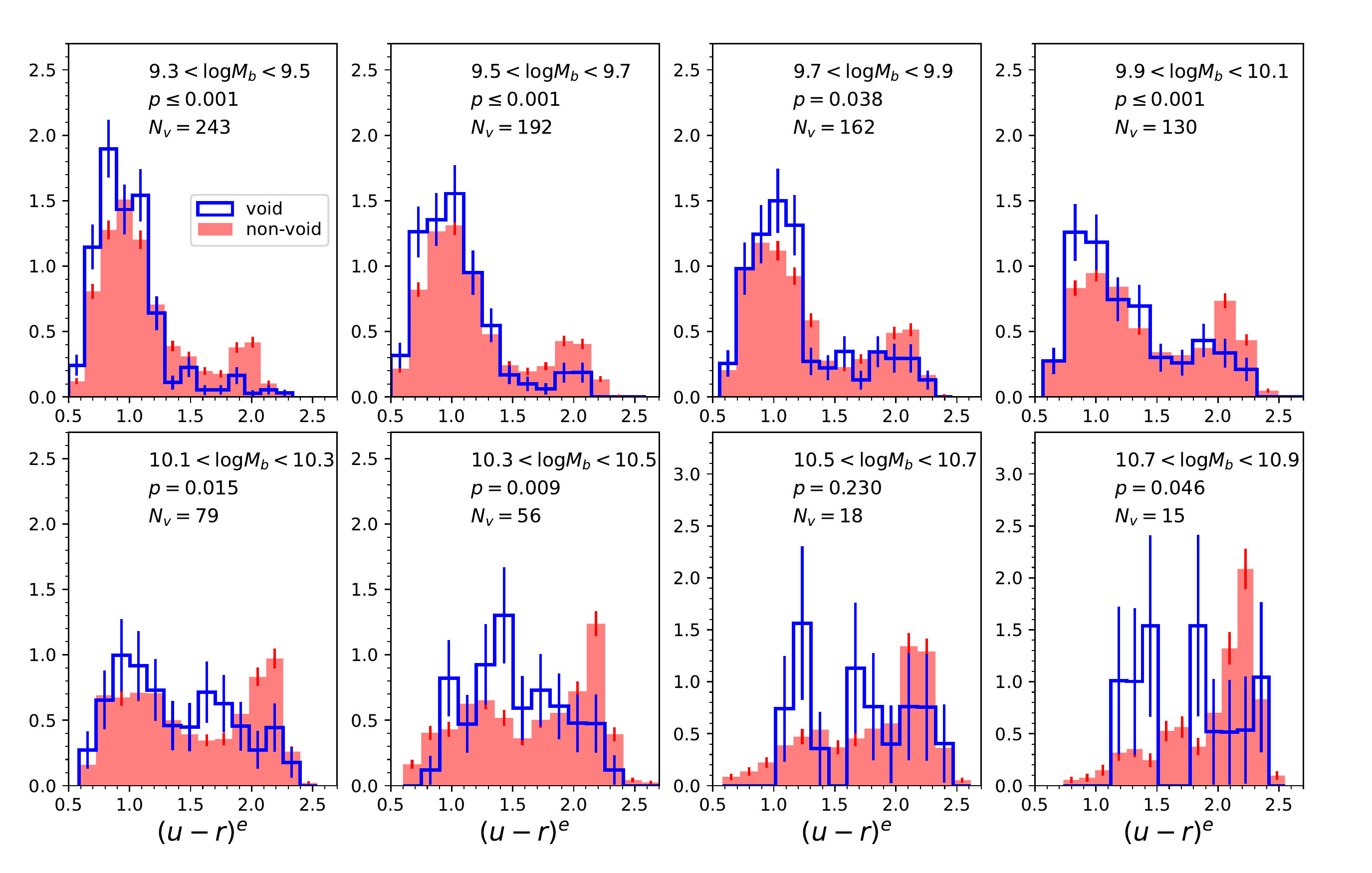}}
\caption{$(u-r)^e$ color distributions of void galaxies (blue) and non-void galaxies (red) in eight bins of baryonic mass. The mass bins are the same eight as in Figure \ref{average_prop}. The number of void galaxies inside each mass bin is also shown on each panel, as well as the $p$-value of the AD test statistic between void and non-void galaxies.}
\label{hist_color}
\end{center}
\end{figure*}

\begin{figure*}[htbp!]
\begin{center}
\centerline{\includegraphics[scale=0.65]{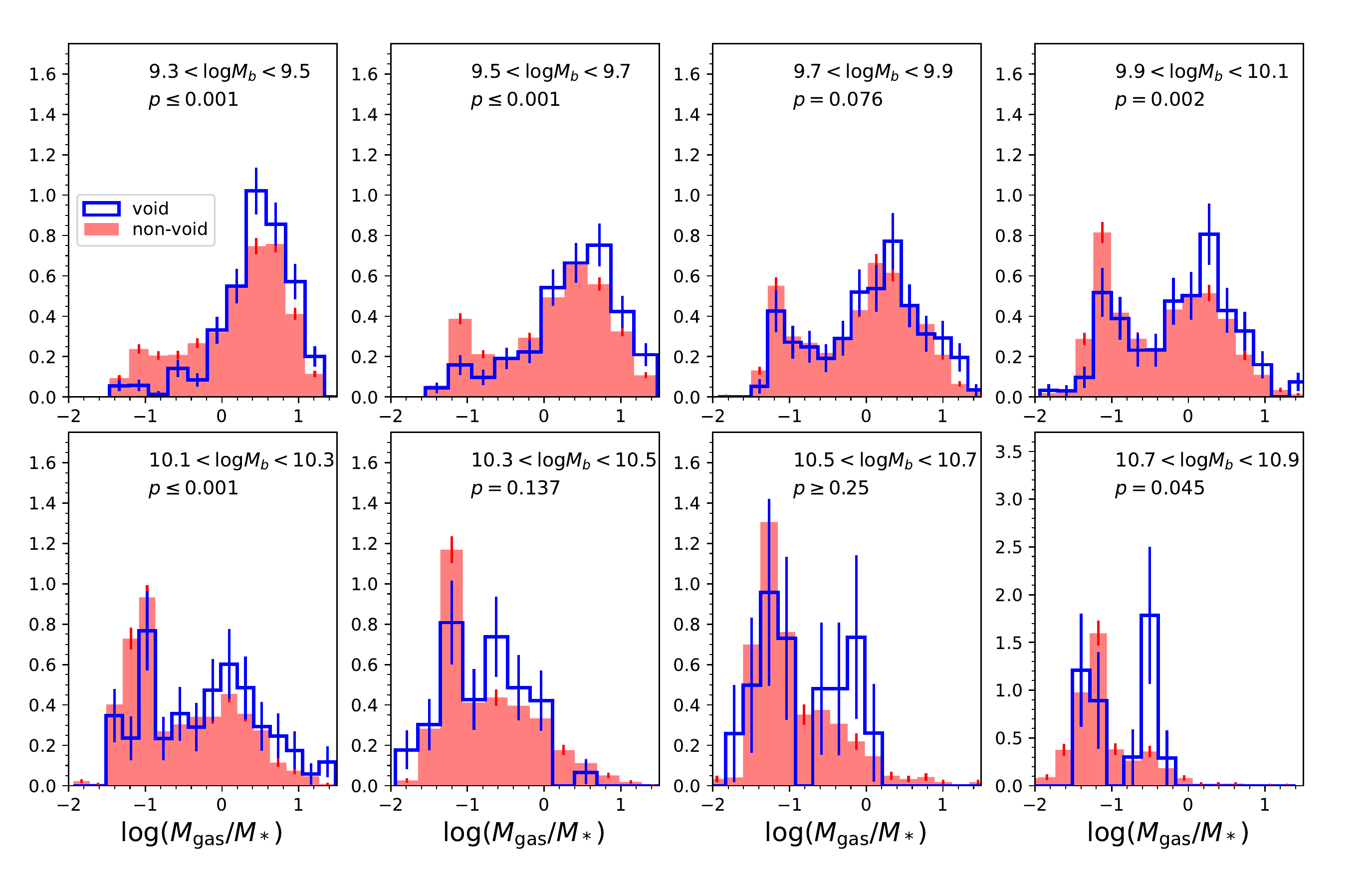}}
\caption{Gas-to-stellar mass ratio distribution of void galaxies (blue) and non-void galaxies (red) in the same eight bins of baryonic mass as in Figure~\ref{hist_color}. As in Figure \ref{hist_color}, we show the $p$-value of the AD test statistic in each panel.}
\label{gs_bins}
\end{center}
\end{figure*}

In order to shed light on why void and non-void galaxies have different properties at fixed mass, we examine some of the property distributions inside each of the baryonic mass bins from Figure \ref{average_prop}. In Figures \ref{hist_color} and \ref{gs_bins}, we examine the color and gas fractions of void and non-void galaxies inside different bins of baryonic mass. We make clear that the samples here are not mass-matched as was done for Figure \ref{new_pdf}, given the narrow range of mass ($\Delta \log(M_b / M_{\odot}) = 0.2$) in which the void and non-void galaxy properties are being compared for these figures. We perform an AD $k$-sample test between the void and non-void galaxy population inside each bin of mass in Figures \ref{hist_color} and \ref{gs_bins} and present the $p$-value in every panel. We remind the reader that the AD $p$-values are capped below values of 0.001 and above values of 0.25 in {\tt scipy.stats}. Although we report a high ($> 0.05$) $p$-value in the panel of Figure \ref{hist_color} where $\log(M_b / M_{\odot}) = 10.5-10.7$, and the panels of Figure \ref{gs_bins} where $\log(M_b / M_{\odot}) = 9.7-9.9$, $10.3 - 10.5$, and $10.5-10.7$ we note that sample sizes can affect the reported $p$-values of these comparisons.

When we investigate the color distribution of void and non-void galaxies inside different bins of baryonic mass (see Figure \ref{hist_color}), we see that at low masses there appears to be a lack of red sequence galaxies inside of voids. Although a red sequence begins to emerge at higher masses, the voids consistently host a slightly bluer population of galaxies and appear to have a deficit in red sequence galaxies when compared to the non-void population. The number of void galaxies begins to drop significantly past $\log(M_b / M_{\odot}) > 10.5$, thus reducing the statistical significance of these results, especially at the highest masses. A similar result holds for gas mass fraction (see Figure \ref{gs_bins}), where void galaxies show a significantly weaker gas-poor peak and a correspondingly stronger gas-rich peak than non-void galaxies. We note that the comparison of color for void and non-void galaxies in Figure \ref{hist_color} agrees qualitatively with Figure 4 of \cite{2012MNRAS.426.3041H}, which shows that void galaxies are bluer than non-void galaxies at fixed magnitude.

\subsection{Are Void Galaxies Different at Fixed Morphology?}

\begin{figure*}[htbp!]
\begin{center}
\includegraphics[scale=0.68]{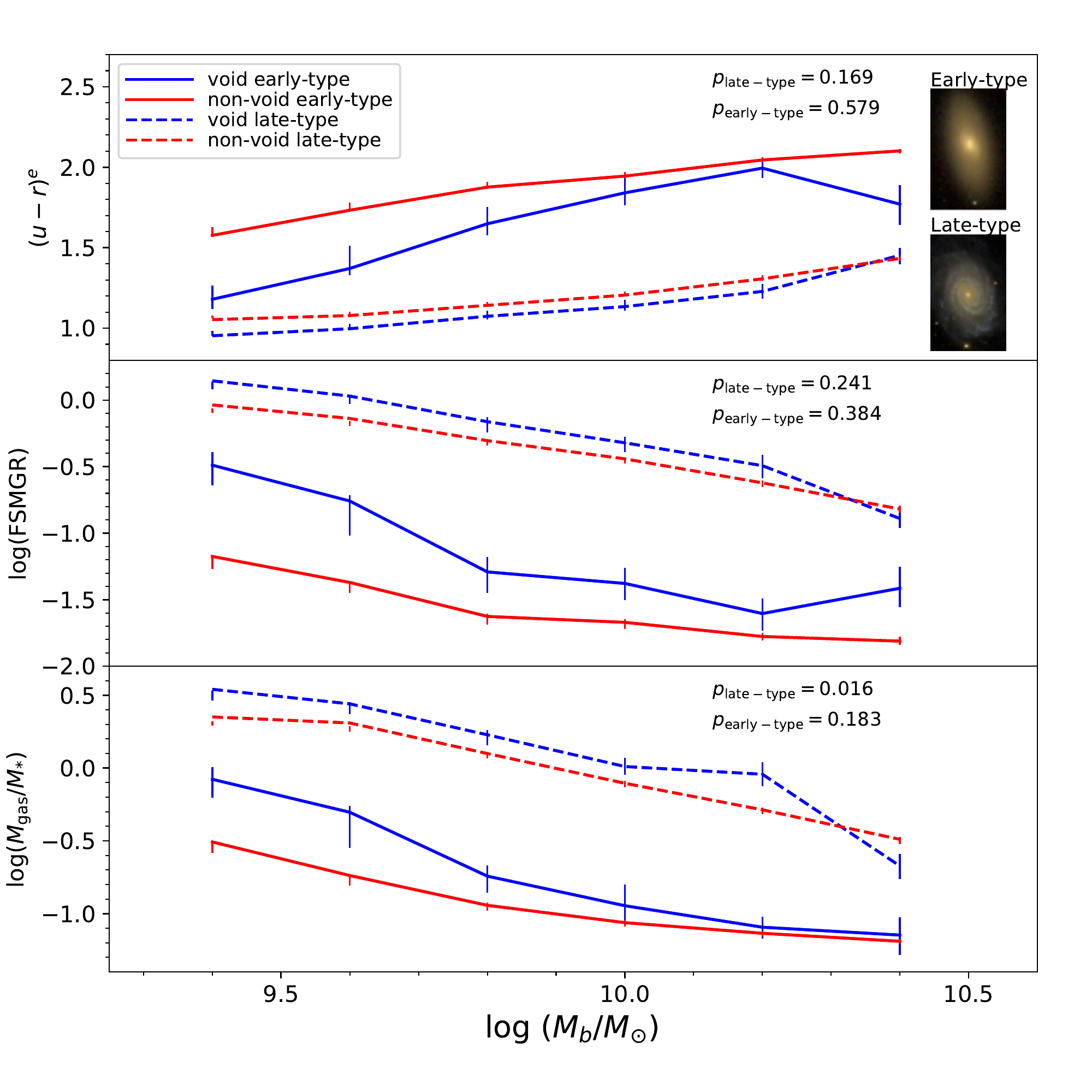}
\caption{Similar to Figure \ref{average_prop}, we show the mean of $(u-r)^e$, $\log (FSMGR)$, and $\log \left(M_{\rm gas} / M_* \right)$ for void (blue) and non-void (red) galaxy late-types (dashed) and early-types (solid) inside six bins of baryonic mass. We also show, as in Figure \ref{average_prop}, the $p$-value of the $T_{n,3}$ test statistic performed between void and non-void galaxies at fixed morphology in each panel.}
\label{average_morph}
\end{center}
\end{figure*}

So far, we have seen that void galaxies are bluer, more star-forming, and more gas-rich than galaxies with higher local densities, and that this is due to more than just the lower mass distribution in voids. In this section we additionally control for morphology in order to determine whether the differences we see between void and non-void galaxies inside a single bin of baryonic mass can be attributed to differences in the morphological distributions. Because voids tend to be occupied by late-type galaxies with lower masses than higher-density regions, we control for morphology in order to determine if the differences we see in color, FSMGR, and gas fraction can be attributed to the larger number of bluer late-types that exist in low-density regions.

In Table \ref{table} we list the number of late types and early types in our void and non-void samples above the baryonic mass completeness limit, as well as the percentage of the total population for each sample.
We note that 83$\%$ of void galaxies are late-types, whereas 73$\%$ of the non-void galaxies are late-types. One might normally expect to find more early-type galaxies inside the non-void environments, however, ECO and RESOLVE are both dominated by dwarf galaxies and late-type fractions are universally higher at low galaxy mass \citep{2006MNRAS.373.1389C}. We remind the reader that ECO and RESOLVE were both designed to explore the low-mass galaxy regime with volume limited samples reaching down to mass limits of $\sim 10^9 M_{\odot}$. In addition, our non-void definition also includes galaxies in regions of average density explaining why the non-void sample is still dominated by late-types. Nonetheless, since there is a higher fraction of late-type galaxies inside voids than non-voids, it is interesting to control for morphology to determine if the differences we see between the void and non-void galaxies can be attributed to the difference in morphological distributions.

\begin{table}[ht!]
\caption{Number of late-type and early-type galaxies in the ECO void and non-void galaxy samples.}
\begin{center}
\begin{tabular}{| c | c | c |}
\hline
Sample & Total N & \% of total\\
\hline
\hline
Void Late-types & 745 & 7.9\%\\
Non-void Late-types & 6296 & 66.6\%\\
Void Early-types & 148 & 1.6\%\\
Non-void Early-types & 2262 & 23.9\%\\
\hline
\end{tabular}
\label{table}
\end{center}
\end{table}

In Figure \ref{average_morph}, we control for both morphology and mass to determine if the void galaxies are intrinsically different from the non-void galaxies. We find that, even at fixed mass and morphology, void galaxies are different from their non-void counterparts. The late-types appear very similar to each other, but void late-types are still bluer, more gas-rich, and more star forming than non-void late-types. Early-type galaxies, however, show much larger variations in color, FSMGR, and the gas fraction between the void and non-void populations at fixed baryonic mass, suggesting that the properties of early-type galaxies correlate far more with environment than the properties of late-type galaxies. 

As described for Figure \ref{average_prop}, we perform a test for equal distributions using the $T_{n,3}$ statistic for void and non-void galaxy late-types and early-types and report the $p$-value in each panel of Figure \ref{average_morph}. We find high $p$-values ($p > 0.05$) for most of these tests, indicating that these comparisons may not be statistically significant. However, we note again that sample sizes could play a role, especially in the case of early-type galaxies as there are only 148 total void early-type galaxies, and 2,262 non-void early-type galaxies.

Our results show that, although morphology correlates highly with color, star formation, and gas-to-stellar mass ratio, environment still plays a role in the observed properties of galaxies at fixed mass and morphology. We controlled for the mass of late-types and still found discrepancies between these properties, even at higher masses where the error introduced by using a predicted baryonic mass diminishes. The early-type population of void galaxies appears to differ significantly from that of the non-void population at fixed mass. This result is consistent with previous results suggesting that early-types in low-density environments evolve differently from the massive red and dead early-types commonly found in clusters, with low-density environments favoring bluer, lower-mass early-types with recent star formation driven by gas-rich mergers or gas inflows \citep{2009AJ....138..579K,2013ApJ...769...82S,2015ApJ...812...89M,2016MNRAS.461.2559L}.


\section{Comparison with Theory: Testing Assembly Bias}

\begin{figure*}[htbp!]
\begin{center}
\centerline{\includegraphics[scale=0.7]{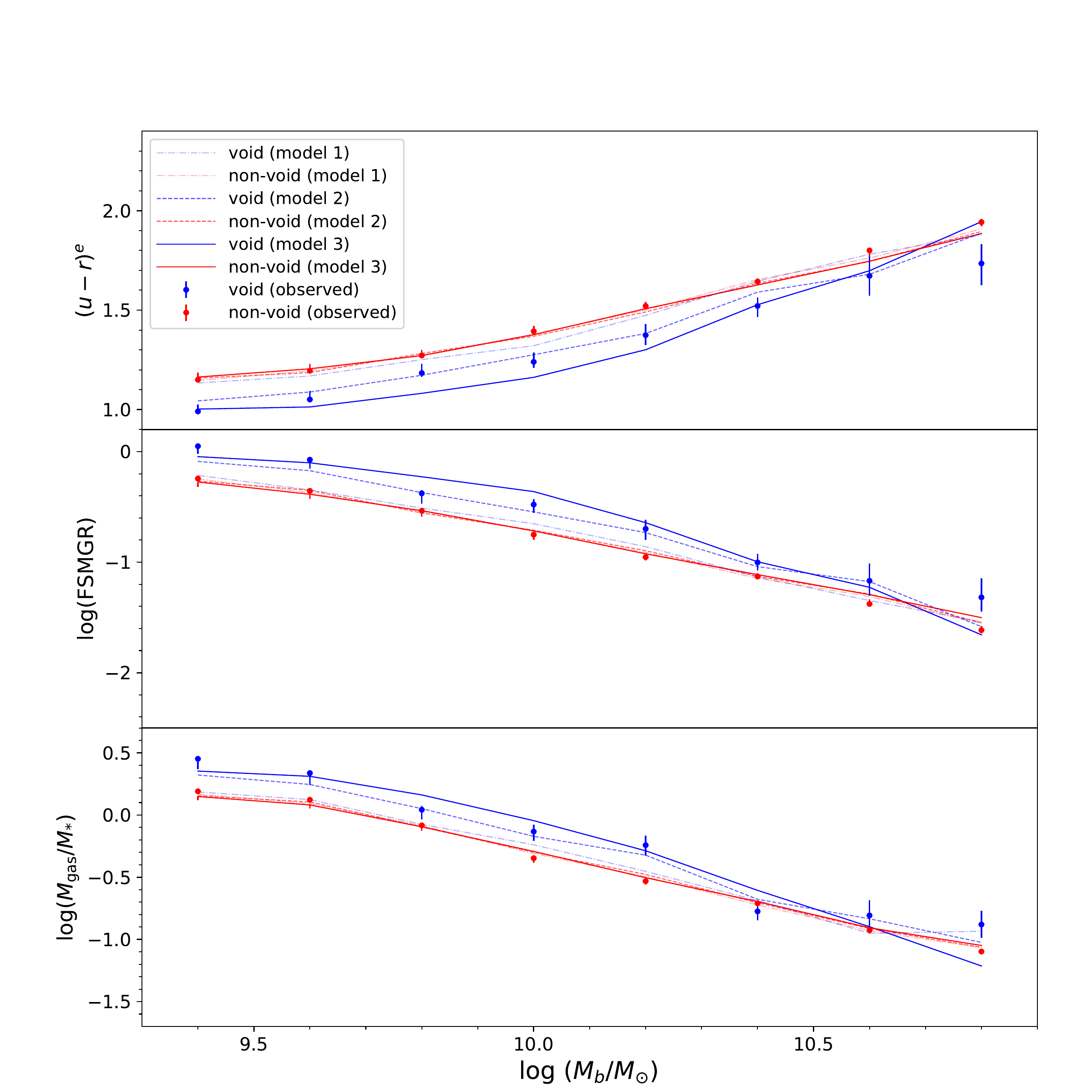}}
\caption{Similar to  \ref{average_prop} and \ref{average_morph}, we show the mean of different observed galaxy properties at fixed baryonic mass in ECO for void (blue circles) and non-void (red circles) galaxies, as well as our results for three mock catalogs (solid, dashed, and dot-dashed lines) that vary by the amount of scatter assigned to the relation between color and the 100 Myr accretion rate of the (sub)halo. Blue lines correspond to the mock void galaxies and red lines correspond to mock non-void galaxies. Model 1 (dot-dashed) assumes no relation between color and the 100 Myr accretion rate. Model 2 (dashed) assigns a correlation coefficient of -0.6 to the relation between color and the 100 Myr accretion rate. Model 3 (solid) assumes a monotonic relation between $u-r$ and the 100 Myr accretion rate of the halo, so no scatter is included in the color assignment of this model.}
\label{mock}
\end{center}
\end{figure*}

While we have shown that void and non-void galaxies exhibit differences in color, FSMGR, and gas fractions, even at fixed mass and morphology, we have not discussed how these differences originated. It is currently unclear whether galaxy properties are primarily affected by halo mass, halo mass $\&$ age, and/or environment. By comparing our results with simulations, we hope to shed light on the relationship between galaxy properties, halo properties, and environment. In particular, we wish to determine whether our results can be attributed to galaxy assembly bias. We do this by constructing mock galaxy catalogs with built-in assembly bias and we analyze them the same way as we have analyzed the ECO sample.

\subsection{Conditional Abundance Matching}

Conditional abundance matching \citep[CAM;][]{2013MNRAS.435.1313H}, is an extension of the subhalo abundance matching (SHAM) method \citep{2004ApJ...609...35K, 2004MNRAS.353..189V, 2006ApJ...647..201C} for assigning observed galaxy properties to dark matter halo properties. While SHAM only connects a single mass-like galaxy property, like stellar mass, to a single mass-like halo property, like maximum circular velocity, CAM allows for two or more halo properties to influence how galaxy properties are assigned in the SHAM algorithm. For example, CAM has been used to study the dependence of star-formation rate on halo formation history \citep[e.g.][]{2013MNRAS.435.1313H, 2014MNRAS.444..729H, 2015MNRAS.446..651W, 2016MNRAS.460.1457S, 2015MNRAS.454.3030P}. Here we briefly describe our implementation of CAM.  

For our fiducial mocks we assign baryonic mass to (sub)halos based on the value of $V_{\rm peak}$, where $V_{\rm peak}$ is defined as the maximum value that the maximum circular velocity, $V_{\rm max}$, attains during the entire assembly history of the (sub)halo.  We define $V_{\rm max} \equiv \text{Max} \left\{ \sqrt{G M( < r)/r} \right\}$, where $M(< r)$ is the mass of the halo enclosed inside some radius $r$, as the maximum circular velocity of a test particle orbiting the halo's gravitational potential well.  It is well established that using $V_{\rm peak}$ for matching in SHAM based methods results in galaxy clustering statistics in closer agreement with observations compared to other (sub)halo properties \citep[e.g.][]{2013ApJ...771...30R, 2017ApJ...834...37L}, although see \citet{2018MNRAS.477..359C} for some caveats.
While the simplest form of SHAM assumes a monotonic relationship between galaxy mass and $V_{\rm peak}$, there is significant evidence for some scatter, $\sim 0.1 - 0.3$ dex in stellar mass at fixed halo mass \citep[e.g.,][]{2009MNRAS.392..801M, 2009ApJ...693..830Y}.  Given this, we add a constant level of scatter of $\sim 0.15$ dex at fixed $V_{\rm peak}$ using the deconvolution method of \citet{2010ApJ...717..379B} with a publicly available implementation\footnote{https://bitbucket.org/yymao/abundancematching}.   

Once baryonic mass is assigned, in order to assign galaxy color, mock galaxies are probabilistically binned in 0.1 dex bins of baryonic mass using the {\tt utils.fuzzy\_digitize} function in {\tt Halotools} \citep{2017AJ....154..190H}. Within each bin, the observed $u-r$ colors of ECO galaxies are used to define a probability distribution: P$_{\text{ECO}}(u-r|M_b)$. If there are $N$ galaxies inside a baryonic mass bin $M_b$, then the galaxy colors are drawn randomly $N$ times from the distribution P$_{\text{ECO}}(u-r|M_b)$ and the draws are rank-ordered. The (sub)halos inside each bin of baryonic mass are also rank-ordered by an age parameter, the 100 Myr accretion rate (i.e. the mass that the halo has accreted in the previous 100 Myrs).  The $u-r$ colors are then assigned to mock galaxies such that redder colors are tied to galaxies that live in (sub)halos with higher values of the 100 Myr accretion rate. The algorithm that assigns colors to mock galaxies also assumes a monotonic relation between $u-r$ and the 100 Myr accretion rate at fixed baryonic mass, so it is necessary to introduce some scatter.  We introduce scatter by degrading the rank order correlation following the method described in appendix D in \citet{2018MNRAS.477..359C}.  

We create eleven mocks with varying levels of scatter between the $u-r$ and 100 Myr accretion rate relation. One mock assumes a correlation coefficient of 0.0, meaning that for that mock the $u-r$ color is not tied to the 100 Myr accretion rate. The other ten mocks have correlation coefficients ranging from -0.1 to -1.0, in steps of 0.1, assigned to the relation between $u-r$ color and 100 Myr accretion rate. The mock with a correlation coefficient of -1.0 assumes a monotonic relation between $u-r$ and the 100 Myr accretion rate. By tying galaxy color to a secondary halo property other than mass, we are explicitly building galaxy assembly bias into our mocks. Moreover, the correlation coefficient for this relation encodes the strength of assembly bias, with a value of 0.0 corresponding to no assembly bias and a value of -1.0 corresponding to maximum assembly bias. As we will show, the amount of scatter that works best for this model to match the data adopts a correlation coefficient of $\rho \sim -0.6$ between the $u-r$ and 100 Myr accretion rate parameters (see Fig.~\ref{mock}). 

Once baryonic mass and $u-r$ color have been assigned to a (sub)halo, we assign other galaxy properties, such as the gas fraction and FSMGR, by identifying an ECO galaxy with similar $M_b$ and $u-r$ as the halo and adopting its other properties. This process results in a mock galaxy distribution that preserves the joint distribution of galaxy properties.

We also experiment with different matching parameters in the models by abundance matching different galaxy properties to $V_{\rm peak}$ and the 100 Myr accretion rate of the dark matter halos, we discuss these results towards the end of this section. 

\subsection{Mock Catalog Construction}

We use the Vishnu cosmological N-body simulation that adopts the has a $130 h^{-1}$Mpc box size and a dark matter particle mass of $3.215\times10^7 h^{-1}M_\odot$ (see \citealt{2019MNRAS.486.1156J} for a detailed description of the simulation and halo catalogs), to construct our mock catalogs. We populate the simulation using the CAM mock population framework in {\tt Halotools} \citep{2017AJ....154..190H}, as described above. We then extract four independent mock ECO galaxy catalogs from the 130 Mpc/h simulation box. For each mock galaxy catalog, we choose a new location for the observer and we place galaxies in redshift space by using the positions and peculiar velocities of their (sub)halo with respect to the observer. We then make identical geometrical cuts in right ascension, declination, and redshift as those in the ECO sample in redshift space, in order to preserve the same geometry as the ECO sample. We make sure to separate the four mock catalogs from each other by at least 10 Mpc/h, thus making them truly independent.

As a result of this procedure, each of the four mock catalogs has the same geometry, similar number density of galaxies, similar overall clustering, and the same joint distributions of observed properties as the ECO sample. Furthermore, galaxy baryonic mass directly correlates with dark matter halo mass, color directly correlates with halo mass accretion rate at fixed mass, and other observed properties like FSMGR or gas fraction inherit indirect correlations with halo mass and accretion rate through their correlations with baryonic mass and color.

It is worth mentioning here that the size of our simulation box may not be large enough to properly model the formation of large cosmic voids. \cite{2017MNRAS.467.4067N} demonstrate that the ``potential fluctuations associated with voids extend over scales of 200-300 Mpc/h." Although we have not shown what impact this consideration could have on our results, we believe it is safe to ignore as our void galaxies are defined by local density rather than large-scale structure. Furthermore, the simulation box is large enough to carve out four independent ECO mock catalogs of densities similar to that of the real ECO catalog, so we believe the simulation box is more than adequate for this type of study.

\subsection{Results}

In Figure \ref{mock}, we show our models with different varying levels of scatter between halo accretion rate and galaxy color at fixed baryonic mass and include the ECO data points from Figure~\ref{average_prop}. We show the results for three models in particular: model 1 which assumes a correlation coefficient of 0.0 to the relation between color and the 100 Myr accretion rate at fixed baryonic mass, model 2 which assumes a correlation coefficient of -0.6 between color and the accretion rate at fixed baryonic mass, and model 3 which assumes a monotonic relation between color and accretion rate (i.e., correlation coefficient is -1.0) at fixed baryonic mass. We find that model 2 matches the data fairly well for both the non-void and void galaxies. Although the mocks are constructed such that the colors of void and non-void galaxies are tied to the halo accretion rate at fixed mass, it is interesting to see in the lower two panels of Figure \ref{mock} that the observed FSMGR and gas fractions trends at fixed mass also hold up in the mock catalogs, which is attributed to the high correlation that exists between color, FSMGR, and gas fractions. The model with a correlation coefficient of 0.0 represents a simple abundance matching of baryonic mass to $V_{\rm peak}$ and assumes no additional correlation between color and the 100 Myr accretion rate (i.e. no galaxy assembly bias). This model cannot reproduce the observed differences between void and non-void ECO galaxy properties. Galaxy assembly bias is thus required to explain the ECO results, at least in the family of models we consider.

Our resulting mock catalogs also do a fairly good job of matching the data when we abundance match stellar mass to $V_{\rm peak}$ and correlate $M_{\rm gas}/M_{*}$ to the 100 Myr accretion. When we abundance match FSMGR to the 100 Myr accretion rate, the model matches the data, but only when the correlation is weak. It should not be surprising that the gas fraction and $(u-r)^e$ parameters both work in the model when matched to the 100 Myr accretion rate because the $M_{\rm gas}/M_*$ for this sample is predicted from color via the PGF technique. The models fail, at all variations with scatter, when we match luminosity to $V_{\rm peak}$ and/or match FSMGR to the 100 Myr accretion rate. While we cannot confirm that the evolution of galaxies is determined solely by assembly bias, as we do not analyze simulations that include the baryonic physics present in different environments, we do show that the void and non-void galaxy trends at fixed mass are recovered in mock catalogs constructed through conditional abundance matching when certain galaxy parameters are abundance matched to the right (sub)halo mass and age parameters.

\section{Discussion}
\subsection{Comparison with Previous Results}
 Our results are in general agreement with observational results that have found enhanced specific star formation rates in underdense regions \citep{2004ApJ...617...50R,1999AJ....118.2561G,2000AJ....119...32G,2007ApJ...654..702M}. More recently, \cite{2014MNRAS.445.4045R} showed that specific star formation rates are higher in void galaxies, and that this is due to more than just the difference in mass distributions amongst different environments. Our results are also in agreement with those of \cite{2012MNRAS.426.3041H} who find that void galaxies are statistically bluer than non-void galaxies, even at fixed magnitude.

In comparison with theory, \cite{2006MNRAS.372.1710P} use a semi-analytic model to study the ratio of blue and red galaxies in void and non-void regions, where the void sample is selected from maximal, non-overlapping spheres with a radius larger than $10 \, h^{-1}$ Mpc and devoid of bright of galaxies. The results of their simulation agree with their observed results: the color distribution of galaxies is bimodal in both void and non-void regions, however, the ratio of blue to red galaxies is higher in voids (although this is more pronounced in the simulation than in the observation). They then control for color and find that the sSFR of void galaxies is similar to that of non-void galaxies in their simulation. They conclude that galaxy properties are not affected by environment at fixed color. 

\cite{2011ApJ...735..132K} used a hydrodynamical simulation to study the environmental dependence of galaxies in voids. They predicted the trends we observe, but only at faint luminosities. Their void galaxy sample is defined by a large underdense region with a diameter of $\sim 30 \, h^{-1}$ Mpc residing at the center of their simulation box. They find that their void galaxies are bluer and more star-forming at fixed luminosity for faint galaxies ($M_r > -16$), but at higher luminosities ($M_r < -18$) they find no difference between the void and non-void galaxies. Although we do not directly control for luminosity, we show at fixed baryonic mass that there is a difference between void and non-void galaxies well above the completeness limit of $M_b = 10^{9.3}$ M$_{\odot}$, and most of these galaxies have luminosities brighter than $M_r = -18$. 

More recent work by \cite{2015ApJ...812..104T} used a hydrodynamical simulation to analyze stellar and halo mass growth in an overdense and underdense region of a simulation box (on scales of $\sim 20$ Mpc). Their simulation produced halos with earlier formation times in overdense regions. First, they found that the stellar-to-halo mass ratio depends on large-scale environment and not just halo mass. They also found that the specific star formation of $z=0$ galaxies is higher in the underdense region of their simulation box when compared to galaxies in the overdense region. Their work suggests that environment, in addition to halo mass, should be considered when assigning stellar masses to halos.

\subsection{E/S0s as Merger Remnants in Voids}
We examine nearest neighbor distances for void galaxies in order to investigate whether there is a merging population of void galaxies that could explain the presence of void early-types. If we consider that possible mergers occur in galaxies that have a nearest neighbor at $< 100$ kpc, we find that an extremely small fraction ($< 1 \%$) of void galaxies look like candidates for mergers that will later become early-types. Given the larger fraction of void galaxies that are considered early-types, we can interpret this in several ways. One possible explanation could be that the early-types we see do not arise as a consequence of merging or other interactive transformation processes, thus challenging assumptions about how early-types form in the universe. 

Observational and theoretical studies of blue early-types do suggest the likelihood, especially in low-density environments, of formation in gas-rich mergers, but gas accretion may also be important \citep{2008MNRAS.388L..10B,2009AJ....138..579K,2013ApJ...769...82S,2014MNRAS.442.1830V}. Another possible explanation for the low number of merger candidates relative to early-types could be that the timescale on which galaxies live in a pre-merger state is far outlived by the lifetime of the post-merger remnant, consistent with a post-merger disk regrowth scenario \citep{2009AJ....138..579K,2013ApJ...769...82S,2015ApJ...812...89M}. Finally, we note that our analysis includes only galaxies brighter than $M_r = -17.33$ and above the baryonic mass completeness limit, so does not rule out current interactions between the void early-type galaxies in our sample and neighboring galaxies whose masses and luminosities fall below our completeness limit.

\section{Summary}

In this work, we have analyzed a sample of void galaxies defined by an Nth nearest neighbor method using the ECO catalog and also consider the RESOLVE-A semester in order to determine the impact of using predicted baryonic masses on the analysis. We examine the color, FSMGR, gas content, and morphology of galaxies in voids and outside of voids.  Our results are as follows.
 
\begin{itemize}
\item We find the void galaxies to have a lower baryonic mass distribution than the non-void galaxies, with an excess of galaxies at $M_b < 10^{9.6} M_{\odot}$ and a deficit of galaxies at mass above $M_b \sim 10^{10} M_{\odot}$ with respect to the non-void galaxy baryonic mass distribution. There hardly exist any void galaxies at $M_b > 10^{11} M_{\odot}$.

\item The void galaxies are bluer, more gas-rich, and more star forming than non-void galaxies, although this is partly driven by differences in the mass distributions of the void and non-void samples. When we mass-match the non-void sample to have an identical baryonic mass distribution as the void sample, these differences are approximately cut in half but persist, showing that environment affects galaxy properties even at fixed mass.

\item When we study the averaged properties in mass bins, we find that void galaxies are bluer, more gas-rich, and more star forming than non-void galaxies at all masses. Satellite galaxies also contribute to the variations we see in the observed mean quantities between void and non-void galaxies at fixed mass, however,  the trends persist (albeit more weakly) when we remove the satellites and only consider central galaxies.

\item Controlling for morphology, we find that the galaxies in voids are bluer, more gas-rich, and more star-forming than their non-void counterparts even at fixed mass and morphology. This is true for both early-type and late-type galaxies, though the differences are significantly smaller for late types.
Incidentally, we also find that voids host a distinctive population of early-types that are bluer and more gas-rich than the typical quenched red-sequence early-types that exist in clusters, consistent with prior work associating low-mass, gas-rich blue E/S0s with low-density and low halo-mass environments \citep[][]{2009AJ....138..579K,2015ApJ...812...89M}.

\item When we construct mock catalogs with built-in galaxy assembly bias using conditional abundance matching, we observe that the color, FSMGR, and $M_{\rm gas}/M_*$ of void and non-void galaxies follow trends similar to those seen in the observed data. We find that the mock catalogs typically match the data to a fair degree when stellar mass or baryonic mass are abundance matched to halo $V_{\rm peak}$ and $(u-r)^e$ or $M_{\rm gas}/M_*$ are abundance matched to the halo mass accretion rate with a modest amount of scatter introduced in the matching. For mock catalogs in which the color or gas fraction is not tied to the age of the (sub)halo, the trends in color, FSMGR, and $M_{\rm gas}/M_*$ for void galaxies are indistinguishable from the non-void galaxies, thus suggesting that galaxy assembly bias is required to explain our observational results.

\end{itemize}

We have, for the first time, performed an analysis on a void and non-void galaxy population while controlling for both mass and morphology simultaneously. Our results suggest that galaxies living in low-density regions are intrinsically different from the rest of the galaxy population, and that this is due to more than just a difference in mass and morphology distributions.

\section*{Acknowledgements}
This project has been supported, in part, by the National Science Foundation (NSF) through Career Award AST-1151650. J.F. acknowledges support from the University of Texas at Austin, NSF grants AST-1413652, AST-1757983, and AST-1614798 and NSF GRFP grant DGE-1610403. A.J.M acknowledges support from the Vanderbilt University Stevenson Postdoctoral Fellowship. Parts of this research were conducted by the Australian Research Council Centre of Excellence for All Sky Astro-physics in 3 Dimensions (ASTRO 3D), through project number CE170100013. The authors acknowledge the Texas Advanced Computing Center (TACC) at The University of Texas at Austin for providing High-Performance Computing (HPC) resources that have contributed to the research results reported within this paper. URL: http://www.tacc.utexas.edu.

This work is based on observations from the SDSS. Funding for SDSS-III has been provided by the Alfred P. Sloan Foundation, the Participating Institutions, the National Science Foundation, and the U.S. Department of Energy Office of Science. The SDSS-III web site is http://www.sdss3.org/. SDSS-III is managed by the Astrophysical Research Consortium for the Participating Institutions of the SDSS-III Collaboration including the University of Arizona, the Brazilian Participation Group, Brookhaven National Laboratory, Carnegie Mellon University, University of Florida, the French Participation Group, the German Participation Group, Harvard University, the Instituto de Astrofisica de Canarias, the Michigan State/Notre Dame/JINA Participation Group, Johns Hopkins University, Lawrence Berkeley National Laboratory, Max Planck Institute for Astrophysics, Max Planck Institute for Extraterrestrial Physics, New Mexico State University, New York University, Ohio State University, Pennsylvania State University, University of Portsmouth, Princeton University, the Spanish Participation Group, University of Tokyo, University of Utah, Vanderbilt University, University of Virginia, University of Washington, and Yale University. This work is based on observations made with the NASA Galaxy Evolution Explorer. GALEX is operated for NASA by the California Institute of Technology under NASA contract NAS5-98034. This publication makes use of data products from the Two Micron All Sky Survey, which is a joint project of the University of Massachusetts and the Infrared Processing and Analysis Center/California Institute of Technology, funded by the National Aeronautics and Space Administration and the National Science Foundation. This work is based in part on data obtained as part of the UKIRT Infrared Deep Sky Survey. This work uses data from the Arecibo observatory. The Arecibo Observatory is operated by SRI International under a cooperative agreement with the National Science Foundation (AST-1100968), and in alliance with Ana G. Mndez-Universidad Metropolitana, and the Universities Space Research Association. This work is based on observations using the Green Bank Telescope. The National Radio Astronomy Observatory is a facility of the National Science Foundation operated under cooperative agreement by Associated Universities, Inc. Swift UVOT was designed and built in collaboration between MSSL, PSU, SwRI, Swales Aerospace, and GSFC, and was launched by NASA. We would like to thank all those involved in the continued operation of UVOT at PSU, MSSL, and GSCF.

\bibliography{void_draft}{}
\bibliographystyle{apj}

\end{document}